\documentclass[%
aps,
reprint,
showpacs,
 amsmath,amssymb,
 aps,
prc,
]{revtex4-2}

\usepackage{graphicx}
\usepackage{dcolumn}
\usepackage{bm}

\begin{document}


\title{$R$-process nuclei in the rare-earth region above $N=126$ in
Relativistic Hartree-Bogoliubov theory: A preponderance of shape-coexistence
in nuclei}
\author{M.M. Sharma}
 \email{madan.sharma@ku.edu.kw}
\affiliation{Physics Department, Kuwait University, Sabah Al Salem University City,
  P.O. Box 2544, Safat 1320, Kuwait}

\date{\today}
\begin{abstract}
The shell structure of nuclei beyond $N=126$ relevant to the third peak
at $ A \approx 195$ in $r$-process nuclear abundances in investigated.
Within the framework of the relativistic
Hartree-Bogoliubov theory in deformed space, a comparative study of the
response of nuclei with three different model Lagrangians NL-SV1, SVI-2 and
NL3 is made. Ground-state properties of even-even nuclei in the isotopic chains 
of $Z=58-72$ in the highly neutron-rich region are investigated. 
A comparison of binding energies with mass formulas shows that in the extreme
region of the $r$-process path, the RMF model SVI-2 without nonlinearities
of $\sigma$ and $\omega$ mesons shows an excellent agreement with the mass
formula HFB-24. It is shown that in all the above Lagrangian models, in the
neutron rich region between the shell closures $N=126$ and $N=184$,
there is a preponderance of nuclides exhibiting coexistence of
well-deformed prolate and oblate shapes 
in the ground state. A prolate ground state is shown to evolve
smoothly to a coexisting oblate ground state  via a near degeneracy
with successive addition of neutrons. Induced by an onset of deformation 
for $N \ge 130$, nuclei approaching the $r$-process path are shown to 
exhibit enhanced stability in many isotopic chains ($Z > 60$). 
This has the effect of shifting the expected neutron drip line 
significantly farther into the neutron-rich region especially
for the higher $Z$ chains. Consequences of the enhanced stability, the shape
co-existence and degeneracy of shapes in nuclei 
on the $r$-process nucleosynthesis are discussed. 
\end{abstract}
\pacs{21.10.Dr, 32.30.Fe, 21.60.Cs, 21.65.+f, 26.30.-k, 27.80.+w, 27.90.+k}
\maketitle


\section{\label{sec:level1}Introduction:\protect}
More than half of nuclei heavier than Fe are synthesized in the $r$-process
nucleosynthesis \cite{BBFH.57,Hille.78,Kratz.93}. 
In this process extremely neutron-rich nuclei far away from the 
stability line are produced in environments of high neutron densities 
and high temperatures. Due to extremely large isospin, the ensuing nuclei 
are highly unstable and undergo a sequence of neutron capture accompanied by 
$\beta$-decays. This gives birth to heavy elements in nature. 
The $r$-process path passes through the magic numbers $N=$~50, 82 and 126 
at different mass numbers. The synthesis of nuclei around these 
magic numbers is manifested in peaks around  $A \sim$ 80, 130 
and 190 in the $r$-process nuclear abundances. 
Due to extremely large neutron excess, nuclei near the $r$-process 
path are experimentally inaccessible, especially those in the heavy mass 
region. Whilst significant progress has been made in understanding 
the $r$-process nucleosynthesis, the site of heavy element synthesis is 
still being debated. Recent observations  of gravitational waves
\cite{GW17.2017, Thiel.2017} have suggested neutron-star mergers as
potential site for nucleosysythesis.

The shell effects in nuclei owe their existence to the presence of magic 
numbers. The strength of the shell effects near the $r$-process path play a 
dominant role in understanding the nucleosynthesis 
of heavy nuclei \cite{Pfeiffer.01}. Quenching of the shell 
effects at $N=82$ near the r-process path was invoked in order to 
reproduce the $r$-process abundances in the second peak
\cite{Pfeiffer.97}. Anomalous behaviour of $2^+$ states in neutron-rich 
Cd isotopes was cited as evidence of the shell quenching
\cite{Dillmann.02}. However, results \cite{Jungclaus.07, Rodriguez.08,
Gorska.09} on level schemes of $^{130}$Cd have shown little evidence of 
quenching at $N=82$. A further analysis \cite{Hwang.08} has shown that 
shape-coexistence in Cd isotopes similar to that found in Hg isotopes 
would explain the larger half-lives and thus enhanced abundances 
without resorting to the shell quenching. These works have confirmed 
the absence of quenching predicted within the RMF theory \cite{SF.02}. 

The shell effects at $N=126$ contribute to nuclear abundances in the third peak
of the $r$-process nucleosynthesis. An interplay between the deformation and
$\beta$-decay rates were shwon to affect the nuclear abuadnces near the third
peak \cite{Surman.97}. Such pictures are based upon simulations. However,
no experimental data are available on nuclei in this
region, which involve extraordinarily large values of isospin.
Hence much of the understanding depends on theoretical predictions. 
Studies of shell effects in nuclei in this extreme region were carried out
earlier. Within the framework of the spherical Relativistic Hartree-Bogoliubov 
(RHB) approach using the Lagrangian model NL-SV1 \cite{SFM.00}, 
shell effects at $N=126$ were shown to remain strong even as 
the neutron drip-line approaches \cite{FS.06}. Nuclei away from the shell
closure pick up deformation rapidly \cite{Guzman.15}. It is imperative
to explore nuclei near the $r$-process path by taking into account
deformation of nuclei.
 
In this work, I have investigated the ground-state properties of nuclei
stretching up to the neutron drip line beyond the shell closure 
$N=126$ within the framework of the Deformed Relativistic 
Hartree-Bogoliubov (DRHB) theory by considering an axially deformed
configuration of nuclei. The DRHB theory is based upon the 
pairing mechanism employing the Bogoliubov 
quasi-particle scheme in a deformed basis.  

\section{\label{sec:level2}The Relativistic Mean-Field Theory: 
Formalism\protect}

The RMF theory and lately called as Covariant Density Functional theory
is based upon the mean-field Lagrangian that describes 
nucleons as Dirac spinors moving in the meson fields \cite{SW.86}:

\begin{eqnarray}
{\cal L}&=& \bar\psi \left( \rlap{/}p - g_\omega\rlap{/}\omega -
g_\rho\rlap{/}\vec\rho\vec\tau - \frac{1}{2}e(1 - \tau_3)\rlap{\,/}A -
g_\sigma\sigma - M_N\right)\psi\nonumber\\
&&+\frac{1}{2}\partial_\mu\sigma\partial^\mu\sigma- 
\frac{1}{2} m^2_\sigma \sigma^2_{} + 
\frac{1}{2}m^2_\omega\omega_\mu\omega^\mu +
\frac{1}{2} m^2_\rho\vec\rho_\mu\vec\rho^\mu\nonumber\\ 
&&-\frac{1}{4}\Omega_{\mu\nu}\Omega^{\mu\nu} -\frac{1}{4}F_{\mu\nu}F^{\mu\nu}
-\frac{1}{4}\vec R_{\mu\nu}\vec R^{\mu\nu}\nonumber\\&&
+~U_{NL}
\label{eq:lagra}
\end{eqnarray} 
where $M_N$ is the bare nucleon mass and $\psi$ is its Dirac spinor. The 
nuclear interaction is produced by exchange of the scalar meson ($\sigma$), 
the isoscalar vector meson ($\omega$) and the isovector vector meson 
($\rho$). The Coulomb interaction between protons is represented by 
the electromagnetic potential $A^\mu$. The meson masses are 
$m_\sigma$, $m_\omega$ and $m_\rho$ and their coupling 
constants are $g_\sigma$, $g_\omega$, $g_\rho$, respectively. The last term
$U_{NL}$ in the Lagrangian density of Eq.~(\ref{eq:lagra}) represents 
nonlinear potentials for $\sigma$ mesons or $\omega$ mesons, or 
both or an admixture of $\sigma$ and $\omega$ meson terms
as discussed in Section~\ref{sec:models}.

The variational principle leads to the Dirac equation:
\begin{equation}
\{ -i{\bf {\alpha}} \nabla + V({\bf r}) + \beta [ m* ] \}
~\psi_{i} = ~\epsilon_{i} \psi_{i}
\end{equation}
where $V({\bf r})$ represents the $vector$ potential:
\begin{equation}
V({\bf r}) = g_{\omega} \omega_{0}({\bf r}) + g_{\rho}\tau_{3} {\bf {\rho}}
_{0}({\bf r}) + e\frac{1-\tau_{3}}{2} {A}_{0}({\bf r})
\end{equation}
and $S({\bf r})$ is the $scalar$ potential
\begin{equation}
S({\bf r}) = g_{\sigma} \sigma({\bf r})
\end{equation}
which defines the effective mass as:
\begin{equation}
m^{\ast}({\bf r}) = m + S({\bf r})
\label{effm}
\end{equation}

The pairing of nucleons is based upon Bogoliubov quasi-particle 
scheme. The eigenstates are eigenvectors of the generalized 
single-particle Hamiltonian containing the self-consistent 
mean-field and a pairing potential $\hat{\Delta}$ which represents 
particle-particle correlations. Within the Hartree approximation for the 
mean-field, the RHB equations are written as
\begin{equation}
\left(\begin{array}{cc} \hat{h}_D-m-\lambda & \hat{\Delta} 
\\ -\hat{\Delta}^* & -\hat{h}_D^*+m+\lambda \end{array}\right)
\left(\begin{array}{r} U \\ V\end{array}\right)_k~=~
E_k\,\left(\begin{array}{r} U \\ V\end{array}\right)_k,
\label{RHB} 
\end{equation}
where $\hat{h}_D$ is the single-particle Dirac Hamiltonian with 
quasi-particle energies $E_k$ and $\lambda$ being the chemical potential.
Here the pairing field 
\begin{equation}
\hat{\Delta}_{ab}({\bf r},{\bf r'})~=~\frac{1}{2}\sum_{cd} V_{abcd}({\bf
 r},{\bf r'}) \kappa_{cd}({\bf r},{\bf r'})
\label{pair}
\end{equation}
is the sum over the matrix elements $V_{abcd}({\bf r}, {\bf r'})$ of a two-body 
pairing interaction. The corresponding pairing tensor is defined by
\begin{equation}
\kappa_{cd}({\bf r},{\bf r'}) = \sum_{E_k > 0}U^*_{ck}({\bf r})V_{dk}({\bf
  r'}).
\label{tensor}
\end{equation}
The RHB equations are solved self-consistently by expanding the fermionic and
mesonic wavefunctions in a basis of 20 harmonic oscillator shells. The nuclei
are assumed to have a cylindrically axial symmetry leading to 
spheroidal shapes. Thus, eigensolutions and eigenvalues 
are obtained in a deformed quasi-particle 
basis. This is then transformed to the canonical basis, which is employed 
to obtain the desired quantities. This approach will henceforth be referred 
to as the Deformed Relativistic Hartree-Bogoliubov (DRHB) approach. Details 
of this have been provided in ref. \cite{Lala.01} and elsewhere. 

For the pairing channel, the Gogny force D1S \cite{Berger.84} is used.
Using the chain of Sn isotopes, the strength of the pairing 
force has been adjusted in order to achieve a good agreement of 
the calculated binding energies with the experimental values.
The superfluid Sn nuclei encompassing a large range of isospin provide 
a good testing ground for calibration of the pairing strength. 

In this work, a comparative study of the ground state properties of nuclei
using three different Lagrangian models each with a different Ansatz is
carried out. These will be discussed below in Section~\ref{sec:models}.

The Lagrangian models that are  employed are the Lagrangian with 
nonlinear vector coupling of $\omega$ mesons NL-SV1, the standard nonlinear
$\sigma$-model with interaction NL3, and the Lagrangian model without 
self-interactions of $\sigma$ and $\omega$ mesons SVI-2. Each of these 
Lagrangian models differs from each other slightly it its attributes. 
A comparative study of Lagrangian models may allow one to see qualitatively
whether the RMF theory has a potential to come closer to predictions of 
macroscopic-microscopic mass formulas that are used widely
for nuclei in extreme regions of astrophysical interest. The
details of these Lagrangian models are discussed below. 

\section{\label{sec:models}The RMF Lagrangian models}

Here a brief description of the formulation of the 
Lagrangian models considered in this work is provided.

\subsection{\label{sec:selfc}The Lagrangian model NL-$\sigma\omega$ 
with nonlinear couplings of $\sigma$- and $\omega$-mesons}

The Lagrangian model NL-$\sigma\omega$ with the addition of a 
quadratic coupling of $\omega$-meson in addition to the standard
nonlinear $\sigma$-meson couplings was developed with a view to
improve description of the shell effects in nuclei along the 
stability line \cite{SFM.00}. Here, the term $U_{NL}$ in 
Eq.~(\ref{eq:lagra}) is
\begin{equation}
\label{nlvsc}
U_{NL} = U_{NL\sigma} + U_{NL\omega}
\end{equation}
where
\begin{equation}
\label{nlsig1}
U_{NL\sigma} = \frac{1}{3}g_2\sigma^3 + \frac{1}{4}g_3\sigma^4. 
\end{equation}
and
\begin{equation}
\label{nlomega}
 U_{NL\omega} = \frac{1}{4}g_4\omega^4. 
\end{equation}
Here, $g_4$ is the quartic coupling constant of the $\omega$ meson.
The Klein-Gordon equations for the meson fields are time-independent
inhomogeneous equations with the nucleon densities as sources.
\begin{equation}
\begin{array}{rl}
(-\Delta + m_{\sigma}^{2})\sigma({\bf r})
             = & -g_{\sigma}\rho_{s}({\bf r})
               -g_{2}\sigma^{2}({\bf r})-g_{3}\sigma^{3}({\bf r})\\         
(-\Delta + m_{\omega}^{2} ) \omega_{0}({\bf r})
             = & g_{\omega}\rho_{v}({\bf r}) + g_4 \omega_0^3({\bf r}) \\  
(-\Delta + m_{\rho}^{2})\rho_{0}({\bf r})
             = & g_{\rho} \rho_{3}({\bf r})\\
-\Delta A_{0}({\bf r}) = & e\rho_{c}({\bf r})
\end{array}
\end{equation}

These equations produce a density dependence of the meson
masses as given by
\begin{equation}
\begin{array}{rl}
\label{vsc}
m_{\sigma}^{*2} = & m_{\sigma}^2 + g_2\sigma 
                   + g_3{\sigma}^2  \\
m_{\omega}^{*2} = & m_{\omega}^2 - g_4{\omega}^2  \\
m_\rho^*    = & m_\rho .
\end{array}
\end{equation}
Thus, both the $\sigma$-meson mass and $\omega$-meson mass exhibit an
implicit density dependence in the $NL$-$\sigma\omega$ model. 

An advantage of the quadratic coupling of $\omega$-meson is a 
softening of the Equation of State (EOS) of nuclear and neutron matter at 
high densities as compared to the non-linear $\sigma$-model \cite{SFM.00}.
This attribute is desirable in order to be able to bring down the maximum
neutron star mass close to the limit observed empirically \cite{ozel.16}. 
The standard non-linear $\sigma$-model is known to overestimate the maximum
neutron star mass significantly. However, in an event of observance of a neutron star mass
significantly higher than 2 solar masses, theories with a stiff EOS may become
admissible, subject to fulfilment of causality.

The interaction NL-SV1 \cite{SFM.00} of the NL-$\sigma\omega$ model is
employed. NL-SV1 was able to reproduce the experimental data away from
the stability line. e.g., the shell effects near the waiting-point nucleus 
$^{80}$Zn \cite{SF.02}. Predictions of the stronger shell effects 
\cite{SF.02} near the $r$-process path using NL-SV1 have been seen 
experimentally \cite{Jungclaus.07,Dworschak.08}. 

\subsection{The Scalar-Vector Interaction model - SVI}
\label{svi}

I have also employed the Scalar-Vector Interaction (SVI) model
in the present work. The SVI model \cite{Sharma.08} 
was developed by the author removing the self interactions of meson fields. 
The idea was to see whether it was still possible to 
describe finite nuclei appropriately without the nonlinearities
of mesonic fields. This attempt was made in Ref. \cite{Sharma.08}. 
It was shown that the nonlinearities could be removed by an 
appropriate combination of couplings of $\sigma$ 
and $\omega$ fields. The self-interaction terms of the mesonic 
fields discussed in Subsections~\ref{sec:selfc} and \ref{sec:sigmac} 
can thus be dispensed with. The introduction of SVI may be 
consistent with a linear realization of the chiral symmetry in a 
suitable scheme \cite{Fomenko.93}. It was shown that the SVI model was
able to provide a good description of the ground-state binding energies 
and charge radii of nuclei \cite{Sharma.08}.

The SVI model consists of the {\it meson-meson interactions}
between $\sigma$ and $\omega$ mesons of the form
\begin{equation}
\label{svieq}
U_{\rm NL} = \frac{1}{2}g_4 \sigma \omega_{\mu}\omega^{\mu}  + 
 \frac{1}{2}g_5 \sigma^2 \omega_{\mu}\omega^{\mu}.
\end{equation}
where $g_4$ and $g_5$ are the coupling constants for
interaction between $\sigma$ and $\omega$ mesons, respectively. 
The Klein-Gordon equations for SVI can be written as:

\begin{equation}
\begin{array}{rl}
\label{svi2}
( -\Delta + m_{\sigma}^{2} ) \sigma({\bf r}) = & -g_{\sigma}\rho_{s}({\bf r})
  + g_4{\omega_0}^2/2 +  g_5\sigma{\omega_0}^2   \\
 ( -\Delta + m_{\omega}^{2} ) \omega_0({\bf r}) = & g_{\omega}\rho_{v}({\bf r})
- g_4 \sigma\omega_0({\bf r})   - g_5 \sigma^2\omega_0 ({\bf r}) \\
 ( -\Delta + m_{\rho}^{2} ) \rho_0({\bf r}) = & g_{\rho} \rho_{3}({\bf r})\\
  -\Delta A_0({\bf r}) = & e\rho_{c}({\bf r}).
\end{array}
\end{equation}
 The effective meson masses are given by:
\begin{equation}
\label{effm3}
\begin{array}{rl}
m_{\sigma}^{*2} = & m_{\sigma}^2  - g_4{\omega_0}^2/(2\sigma) 
               -  g_5{\omega_0}^2 \\
m_{\omega}^{*2} = & m_{\omega}^2 + g_4 \sigma  + g_5 \sigma^2\\
m_\rho^*        = & m_\rho.
\end{array}
\end{equation}
There is an inter-dependence in density dependence of
the $\sigma$ and $\omega$ meson masses on each other. 

\subsection{\label{sec:sigmac}The nonlinear-$\sigma$ model}
The nonlinear sigma (NL-$\sigma$) model is the standard Lagrangian
model that is used most commonly for calculation of the ground-state 
properties of nuclei. An important ingredient of the  (NL-$\sigma$) 
model is the assumption of the nonlinear scalar self-couplings of the 
form \cite{BB.77}:
\begin{equation}
\label{nlsig2}
U_{NL} = \frac{1}{3}g_2\sigma^3 + \frac{1}{4}g_3\sigma^4. 
\end{equation}
Here the parameters $g_2$ and $g_3$ are the nonlinear couplings of the 
$\sigma$-meson in the conventional NL-$\sigma$-model. 

The Klein-Gordon equations for the NL-$\sigma$ model are given by 
\begin{equation}
\begin{array}{rl}
\label{KGSIG}
(-\Delta + m_{\sigma}^{2})\sigma({\bf r})
             = & -g_{\sigma}\rho_{s}({\bf r})
                -g_{2}\sigma^{2}({\bf r})-g_{3}\sigma^{3}({\bf r})\\        
(-\Delta + m_{\omega}^{2}) \omega_{0}({\bf r})
             = & g_{\omega}\rho_{v}({\bf r}) \\  
(-\Delta + m_{\rho}^{2})\rho_{0}({\bf r})
             = & g_{\rho} \rho_{3}({\bf r})\\
  -\Delta A_{0}({\bf r}) = & e\rho_{c}({\bf r})
\end{array}
\end{equation}

These equations produce a density
dependence of the meson masses as given by
\begin{equation}
\begin{array}{rl}
\label{effm2}
m_{\sigma}^{*2} = & m_{\sigma}^2 + g_2\sigma 
                   + g_3{\sigma}^2  \\
m_\omega^*  = & m_\omega \\
m_\rho^*    = & m_\rho .
\end{array}
\end{equation}
Here, only the $\sigma$-meson mass shows an implicit density dependence
in the NL-$\sigma$ model. 

The NL-$\sigma$ model has been a very successful one for 
ground-state properties of nuclei all over the 
periodic table \cite{GRT.90,Ring.96}. The earliest
parameter sets obtained were NL1 \cite{Rein.89}
and NL2 \cite{Lee.86}. These interactions were 
not suitable due to large asymmetry energy $J$ and inordinately large neutron
thickness of nuclei. This problem was resolved in the force
NL-SH \cite{SNR.93} that was developed as one of the very first 
successful parameter sets in the RMF theory. Due to a relatively larger 
value of $K_\infty$ of NL-SH, an improved force NL3 has been 
obtained \cite{Lala.97}. This would allow to get breathing-mode 
Giant Monopole Resonance (GMR) energies in a physically 
acceptable region. NL3 has shown improvements in ground state
properties of nuclei. Several other Lagrangian model have since been 
developed within the $\sigma^3 + \sigma^4$ model \cite{Meng.2010}.

\subsection{The nuclear matter properties and effective mass}
\label{nucmat}

\begin{figure}
\resizebox{0.40\textwidth}{!}{%
   \includegraphics{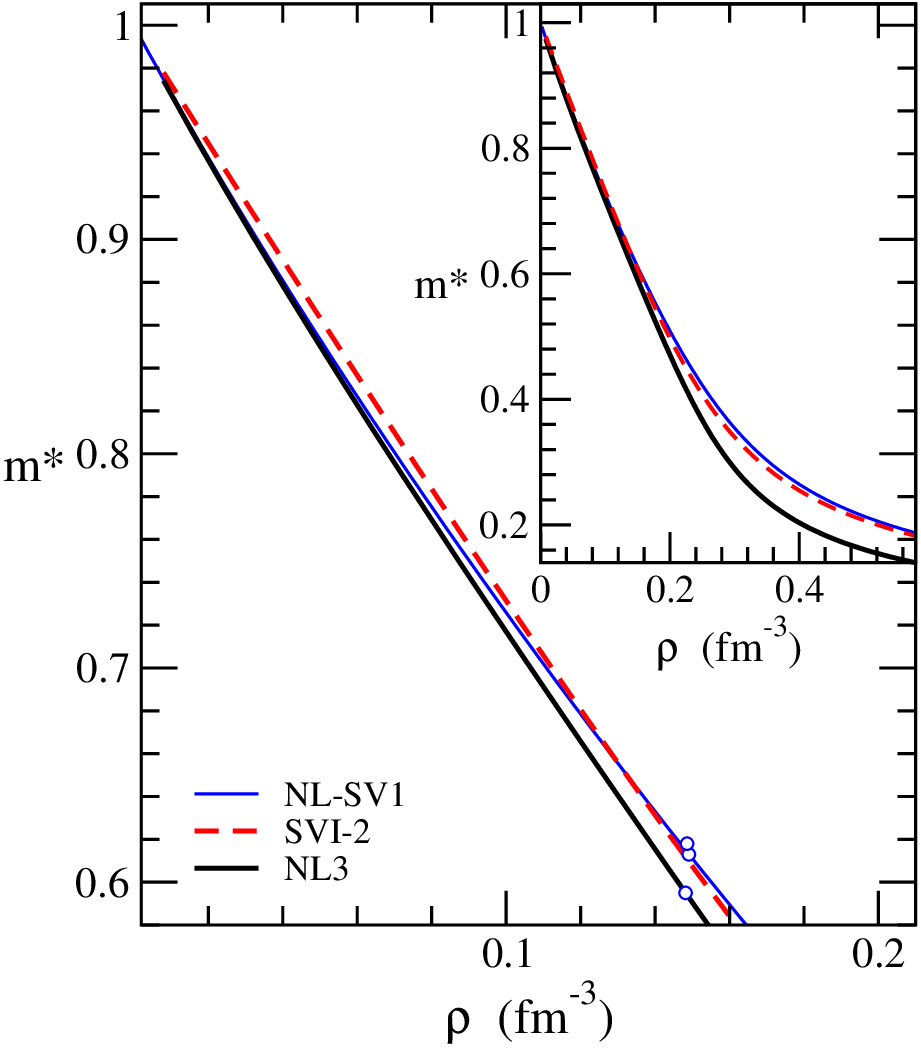}}
\caption{The effective nucleon mass in nuclear matter as a function of nucleon 
density as obtained from the Lagrangian sets NL-SV1, SVI-2 and NL3. The open 
circles denote the saturation point. The inset shows the same the higher
density behaviour.} 
\label{fig:fig1}       
\end{figure}

The nuclear matter properties of successful parameter sets in the 
three Lagrangian models are given in
Table~\ref{tab:table1}. Surprisingly, nuclear matter properties of 
all the three parameter sets, though belonging to three completely 
different Lagrangian models converge around a value, be it
compression modulus $K_\infty$, or binding energy at saturation $E/A$, or
saturation density $\rho_0$, or asymmetry energy $J$. There is a slight 
difference in the effective mass at saturation $m^*$ with the lowest value
$m^* = 0.595$ for NL3 and the largest value being $m^* = 0.621$ for SVI-2.
This difference is about 4\% compared to a 
miniscule difference in other properties amongst the parameter 
sets. It may be noted that all the three parameter
sets reproduce the ground state binding energies of nuclei along the
stability line successfully. It is not accidental that the different
Lagrangian models require very similar values for nuclear matter 
properties in order to be successful for ground state properties of
nuclei. A point may be in order here: Without allusion to the
breathing mode GMR energies of nuclei, the ground state properties
of nuclei fitted by the three Lagrangian models require the compression 
modulus $K_\infty$ to be around 270 MeV. The effective nucleon mass obtained
from the three Lagrangian models are shown in  Fig.~\ref{fig:fig1}.

\begin{table}[h]
\caption{\label{tab:table1}
The nuclear matter properties of the Lagrangian sets in the models 
employed in the present work.} 
\begin{ruledtabular}
\begin{tabular}{l l l c c c c c }
& Model & Set  & $K_\infty$  & $E/A$  & $m^*$ & $\rho_0$  & $J$  \\
&       &       &  (MeV)     & (MeV)  &       & (fm$^{-3}$)& (MeV) \\    
\hline 
& NL-$\sigma\omega$ & NL-SV1  & 270.1 & -16.26 & 0.613 & 0.149 & 37.3 \\   
& NL-$\sigma$    & NL3     & 271.6   & -16.25 & 0.595   & 0.148  & 37.4 \\
 & SVI           & SVI-2   & 271.4   & -16.31 & 0.621   & 0.149  & 36.4 \\
\end{tabular}
\end{ruledtabular}
\end{table}

The density dependence of $\sigma$- and $\omega$-meson masses, 
as given by Eqs. (\ref{vsc}), (\ref{effm2}) and (\ref{effm3}) for 
the three Lagrangian models NL-$\sigma\omega$, NL-$\sigma$ and SVI, 
respectively, are different from one another. The effective nucleon
mass given by Eq.(\ref{effm}) is another variable that depicts to some
extent density dependence of nuclear interaction. I show in 
Fig.~\ref{fig:fig1} the effective nucleon mass in nuclear matter
as obtained with the sets NL-SV1, SVI-2 and NL3 of the 
three Lagrangian models. The low density behaviour of the three
is not much different from each other. As the density increases upto
the saturation density of $\rho_o \approx 0.15$ fm$^{-3}$ 
(indicated by the open circles), the curves
show an increased separation. The Lagrangian model NL3 separates out
whereas NL-SV1 and SVI-2 curves are closer together. The $\sigma$-field
is thus influenced nearly identically by $\omega^4$ coupling in NL-SV1
and couplings between $\sigma$- and $\omega$-fields in SVI-2. The saturation
point for both NL-SV1 and SVI-2 are also very close to each other. 

The inset of Fig.~\ref{fig:fig1} shows the high density behaviour of the
effective mass up to about 4$\rho_o$. NL-SV1 and SVI-2 curves 
are are closer together, where NL3 shows a stronger decrease at higher
densities. This can be correlated to high density behaviour of the
EOS of the nuclear matter. Both NL-SV1 and SVI-2 are known to give
a softer EOS than NL3. A reduced effective mass with NL3 implies 
a stiffer EOS at high densities.

\begin{figure*}[ht]
\resizebox{0.70\textwidth}{!}{%
   \includegraphics{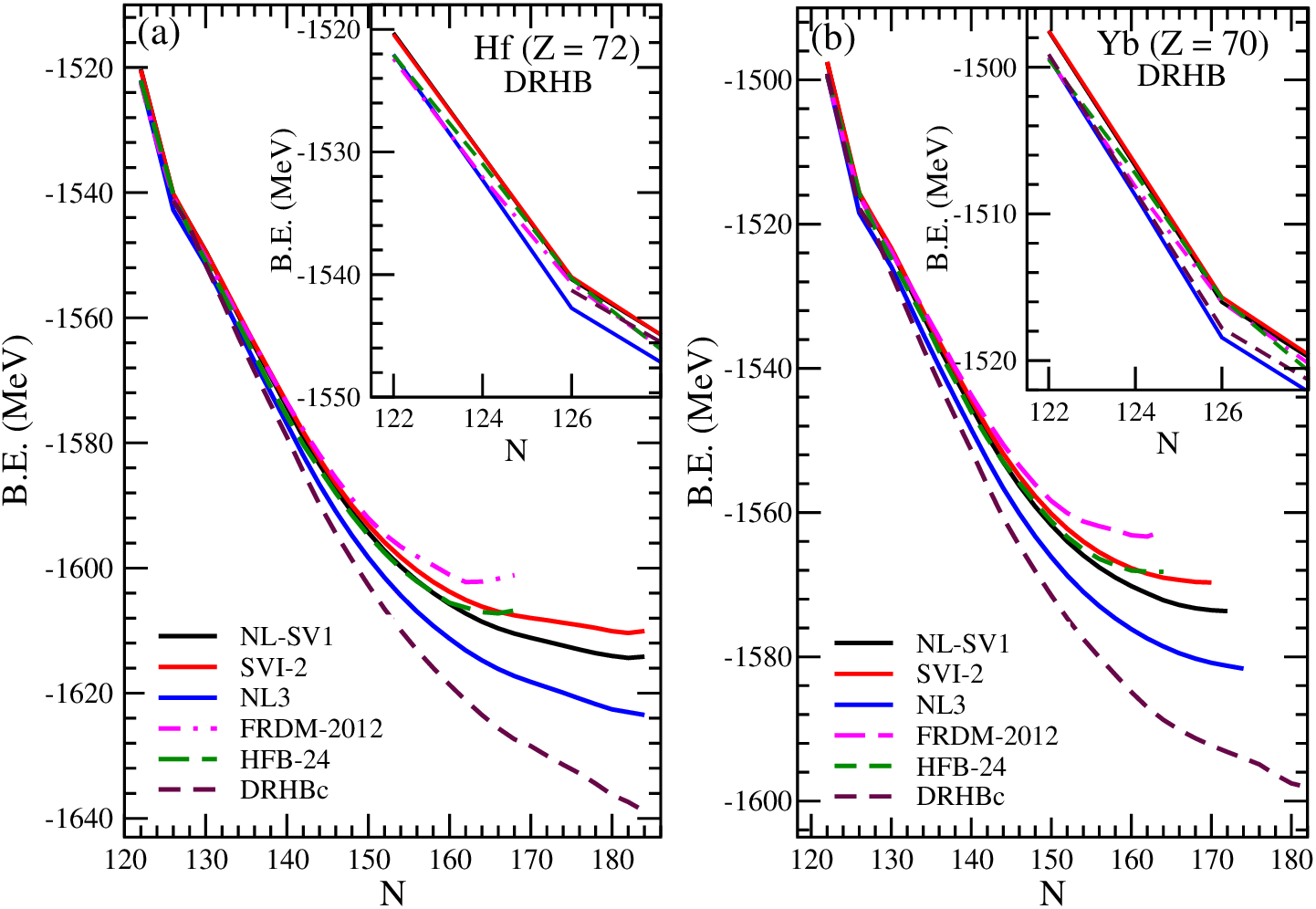}}
\caption{The binding energies of nuclei in (a) Hf and (b) Yb isotopic 
chains obtained with Lagrangian models NL-SV1, SVI-2 and NL3 employing 
the deformed RHB approach. Predictions of mass formulas FRDM-2012 and 
HFB-24 are also shown in comparison. The inset shows the same for nuclei
around $N=126$ depicting the kink at the magic number $N=126$.} 
\label{fig:fig2}       
\end{figure*}
\section{\label{sec:results}Results and discussion}

\subsection{\label{sec:drhb} The DRHB Calculations:\protect}

Here deformed DRHB calculations for even-even nuclei in various 
isotopic chains in the rare-earth region using the three different 
Lagrangian models NL-SV1, NL3 and SVI-2 have been performed.
Ground-state properties  such as binding energies and deformation
properties of even-even nuclides in various isotopic chains ($Z=58-72$)
have been obtained from the magic number $N=126$ up until the arrival
of the neutron drip-line.

\subsection{\label{sec:be}The total binding energy}

The total binding energy of the isotopic chains of Hf ($Z=72$) and Yb ($Z=70$)
obtained with NL-SV1, SVI-2 and  NL3 are shown in Fig.~\ref{fig:fig2}. 
The neutron number spans 
isotopes from just below $N=126$ up until the drip line. Binding 
energies from the mass formula Finite-Range Droplet 
Model - FRDM2012 \cite{Moeller.16} and from the refined  mass
formula HFB-24 \cite{Goriely.13} based upon Skyrme density 
functional fitted to various nuclear properties reaching the 24th version 
are shown for a comparison. The latter is considered to represent the masses
of known nuclei including many new ones in the extreme regions with an
improved accuracy. A comparison is made with the recenlty obtained mass
formula DRHBc \cite{Guo.2024} that is based upon point coupling using the nuclear
interaction PC-PK1 within the Covariant Density Functional Theory. 

Results of NL-SV1 and SVI-2 show good agreement with each other all along
until $N \approx 150$. They show a difference of $\approx$4 MeV in reaching the very 
extreme point of $N=184$. Binding energies from NL-SV1 and SVI-2
show a reasonably good  agreement with those of the mass formulas
from $N=120$ up until $N \approx 150$. The  binding energies for all approaches
show a constant increase as shown by a steep decline of the curves as a 
function of neutron number. The slope of the curves decreases 
above $N = 150$ indicating a smaller contribution to the
total binding energy on further addition of neutrons as nuclei approach
the region of the $r$-process path and eventually reaching the drip line.
Here, the drip line is defined as the point where neutron separation energy
becomes close to zero and the chemical potential reaches near the continuum.
The last point in each curve corresponds to the nucleus
at the drip line. The DRHB results for Hf (Fig.\ref{fig:fig2}(a)) show that
with the RMF models the neutron drip line is extended much farther into
the isospin space with neutron number reaching very close to the next
magic number $N=184$. 

The predictions of mass formulas FRDM-2012 and HFB-24 show agreement 
up until $N=150$; they show differences in the binding energies 
above $N=150$, the difference reaching 4-6 MeV between them 
for $N=168$. FRDM-2012 and HFB-24 have made predictions of masses
for nuclei up to $N \approx 168$ . For extreme nuclei,
FRDM-2012 underbinds nuclei significantly as compared to HFB-24. 

Interestingly, both NL-SV1 and SVI-2 agree very well 
with the predictions of HFB-24 in the extreme region. Flattening of the 
binding energy curve with FRDM-2012 and HFB-24 at $N \approx 160$ shows an onset
of the drip line much earlier than that for the RMF models, impliying
that nuclei in the extreme region with the RMF models are more strongly 
bound than in the mass formulas.

The relative difference between various models becomes clear in the
extreme region of $r$-process path and drip line. Whilst results with
NL-SV1 and SVI-2 are close to each other in much of the region uptil
$N \approx 160$, differences between them start appearing in the extreme region 
beyond $N=160$. Nuclei with NL-SV1 are overbound by $\approx$3-4 MeV as compared
to SVI-2. It is noteworthy that up until $N \approx 150$, both NL-SV1 and SVI-2 show
results in close proximity to those of mass formulas FRDM-12 and HFB-24,
both of which also agree with each other.

Above $N \approx 150$ until $N \approx
160$, NL-SV1 and SVI-2 show binding energies which agree more with HFB-24 than
FRDM. The latter underbinds nuclei by ~4-6 MeV as compared to HFB-24 and Lagrangian
models NL-SV1 and SVI-2. It is remarkable to note that in both the
mass formulas, the drip line arrives much earlier at $N \approx170$, as compared
to that with the RMF models which predict the neutron drip line extended
all the way up until the next magic number $N=184$. Clearly, the RMF models
exhibit much stronger binding for nuclei in the extreme region of the
neutron drip line. Of all the RMF Lagrangian models, NL3 shows the
overbinding for nuclei beginning with $N=126$, that increases rapidly as one
approaches the neutron drip line.

It is instructive to compare the results with those of the mass formula DRHBc.
The latter overbinds nuclei significantly as compared to RMF models and inordinately
as compared to HFB-24 and FRDM-2012. The overbinding increases rapidly as one
moves towards the neutron drip line. For instance, for Hf nucleus with $N=180$, DRHBc
overestimates the binding energy by ~27 MeV as compared to SVI-2, by ~23 MeV
as compared to NL-SV1 and ~16 MeV as compared to NL3.

The inset shows the binding energies for nuclei close to $N=126$. A kink
at $N=126$ reflects the shell effects and magicity due to the magic number
$N=126$. Whilst NL-SV1, NL3 and SVI-2 tend to agree with FRDM-2012 and HFB-24
below $N=126$, NL3 begins diverging with all others as one goes above
$N=126$. The divergence becomes more pronounced above $N=140$. Surprizinlgy,
DRHBc results are closer to those of NL-SV1 and SVI-2.

The trend of binding energies for Yb ($Z=70$) shown in 
Fig. \ref{fig:fig2}(b) is similar to that for Hf ($Z=72$). 
One notable difference is that the drip line has moved down to $N\approx 170$
with SVI-2, to $N \approx 172$ with NL-SV1 and to $N\approx 174$ with NL3,
opposed to that for Hf ($Z=72$). For Hf ($Z=72$), the prolonging of neutron
drip line to the magic number $N=184$ may have been sustained by
strongly binding nature of the magic number. For Yb ($Z=70$) isotopes,
the drip line nuclei are still 10-12 neutrons less than the magic number
$N=184$, as will also be seen in the Section \ref{sec:s2n}

The RMF models NL-SV1 and SVI-2 show a good agreement with each other and
with the mass formulas up until $N \approx 150$ for Yb ($N=70$). NL-SV1 and
SVI-2 show a slight difference of about 2-4 MeV in going to the neutron
drip nuclei. The model NL3 shows the largest divergence amongst all
the RMF models and with the mass formulas. At the drip line, NL3
overbinds the nuclei $~8-9$ MeV as compared to the NL-SV1 and
by $ \approx 12$ MeV as compared to SVI-2. The behaviour of DRHBc is similar to
that seen for the Hf chain, as far as the excessive binding of nuclei is
concerned especially close to the drip line. 

As seen in Fig. \ref{fig:fig2}(a), the mass formulas FRDM-2012 and 
HFB-24 both predict a drip line with 6-8 neutron less than that 
for RMF models. The two mass formulas diverge with each other 
in the total binding energy of about 6 MeV at the drip line, the FRDM-2012
exhibiting a corresponding underbinding.

A comparison of the RMF models with the 
mass formulas shows that the SVI-2 provides the best agreement
with HFB-24. In contrast, the overbinding by DRHBc extends the drip line to
near $N=184$. As in the inset of Fig. \ref{fig:fig2}(a), the inset of
Fig. \ref{fig:fig2}(b) shows that NL3 shows a divergence from the other
RMF models as well 
as with the mass formulas near $N=126$.  However, NL3 and DRHBc results
are closer to each other near $N=126$.

\begin{figure*}[ht]
\resizebox{0.70\textwidth}{!}{%
   \includegraphics{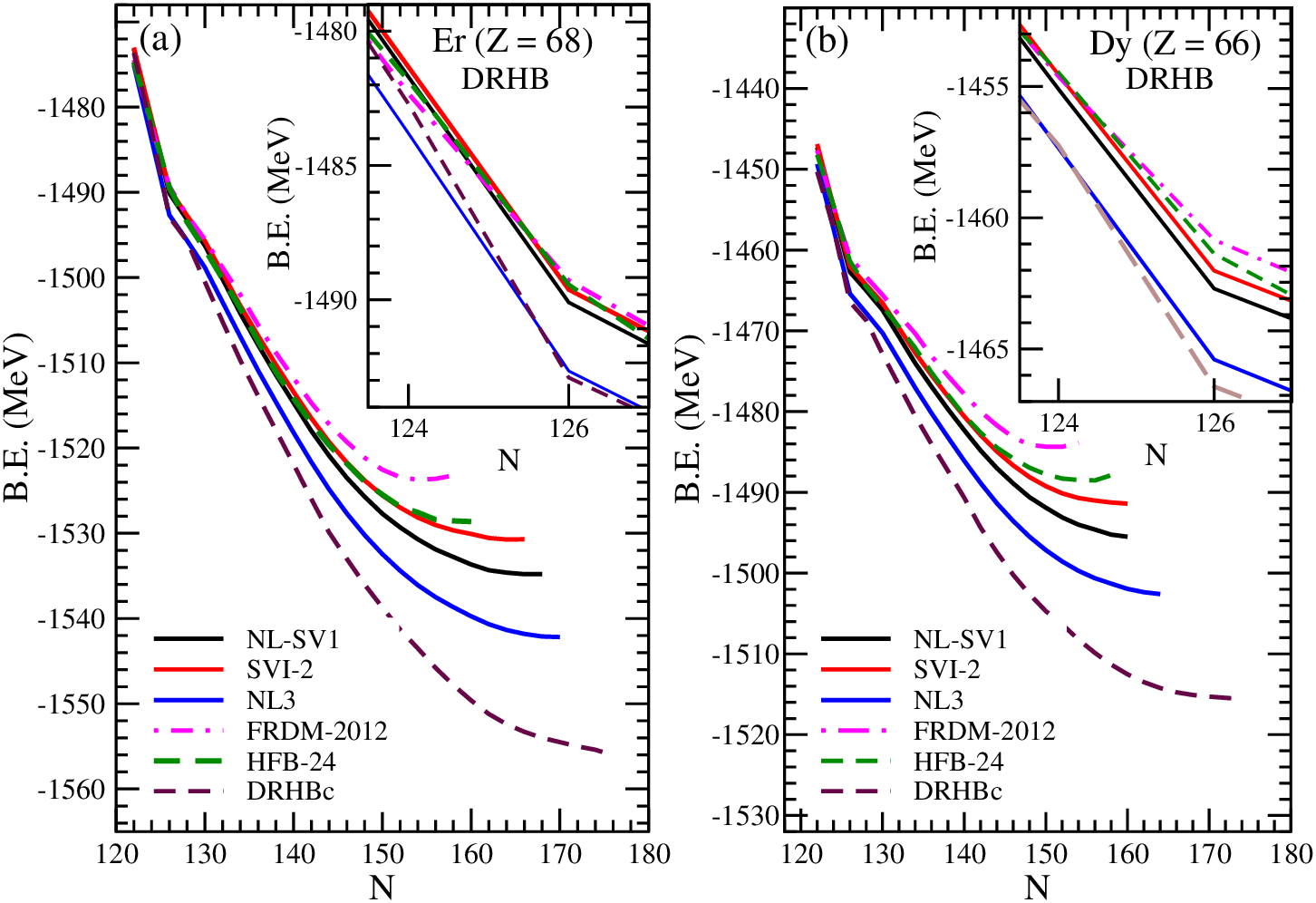}}
\vspace{0.5 cm}
\caption{The same as in Fig. \ref{fig:fig2} for (a) Er and 
(b) Dy isotopic chains.} 
\label{fig:fig3}       
\end{figure*}

Fig. \ref{fig:fig3} shows the binding energies of Er ($Z=68$) and Dy ($Z=66$). 
For Er (Fig. \ref{fig:fig2}(a)), the overall features of the total 
binding energy of nuclei are similar to those of Fig. \ref{fig:fig2}. 
Here the drip line has shifted down to $N \approx 160$. Both the mass formulas 
predict a drip line much closer to that with the RMF models, with 
FRDM-2012 showing it at 2 neutron number less. 

The RMF models NL-SV1 and SVI-2 show again good agreement with each other 
and with the mass formula HFB-24 up to $N \approx 140$. Difference between these
two models start appearing for nuclei above $N \approx 140$ until the approach of drip
line where they differ by 3-4 MeV, the NL-SV1 showing a slight overbinding. 
The model SVI-2 exhibits an excellent agreement with HFB-24 all the way upto
the neutron drip line. NL3, on the other hand, shows a significant divergence 
from both the other RMF models and the mass formulas, beginning with 
$N=126$ all the way up until the neutron drip line. Divergences of NL3
predictions from the mass formula HFB-24 amount to $ \approx 12$ MeV  and 
$ \approx16$ MeV from FRDM-2012. DRHBc overbinds nuclei
excessively in going towards the drip line, as seen for Hf and Yb in 
Fig. \ref{fig:fig2}(a). Consequently, the drip line with DRHBc is pushed
to higher $N$ as compared to all others.

The kink at $N=126$ in the inset of Fig. \ref{fig:fig3}(a) shows that the
shell effects with the RMF models remain strong in conformity with the
predictions of the mass formulas. The divergences of NL3 at $N=126$ and
its neighbourhood are larger than those of insets of Fig. \ref{fig:fig2}.
This is part of a larger trend of NL3 divergences with all other models. On
the other hand, DRHBc agrees well wtih NL3 in the vicinity of $N=126$.

For Dy ($Z=66$) isotopes shown in Fig. \ref{fig:fig3}(b), the overall 
features are very similar to those seen in Fig. \ref{fig:fig3}(a).
The drip line with the RMF models shifts down by 4-6 neutrons as
compared to Er and for the mass formulas it shifts marginally by
2 neutrons. It can be seen that in the extreme region of the neutron
drip line, all the approaches show divergence with each other. However,
SVI-2 shows an impeccable agreement with HFB-24
over a large region until $N \approx 146$. The difference between these two are of
about $2-3$ MeV for nuclei near the drip line. Whilst NL-SV1 and SVI-2 
agree with each other, and the differences between NL-SV1 and SVI-2 are minimal
as the neutron number progresses, these differences grow to $\approx 4$ MeV for
nuclei at the neutron drip line. As seen before, predictions of NL3 diverge
strongly with all the approaches with large discrepancies seen at the drip line.
It is worth noting that there are also systematic differences in predictions of
FRDM-2012 and HFB-24 especially above $N \approx 140$, reaching a difference 
of $ \approx 4-6$ MeV between the two. As always, FRDM-2012 underbinds 
nuclei as compared to HFB-24 and the RMF models. In contrast, DRHBc overbinds
nuclei excessively as was seen for Hf, Yb and Er chains, with the consequence
that the drip line is pushed to much higher $N$ value as compared to all other
approaches.  

\vspace {0.5 cm}
\begin{figure*}[ht]
\resizebox{0.70\textwidth}{!}{%
   \includegraphics{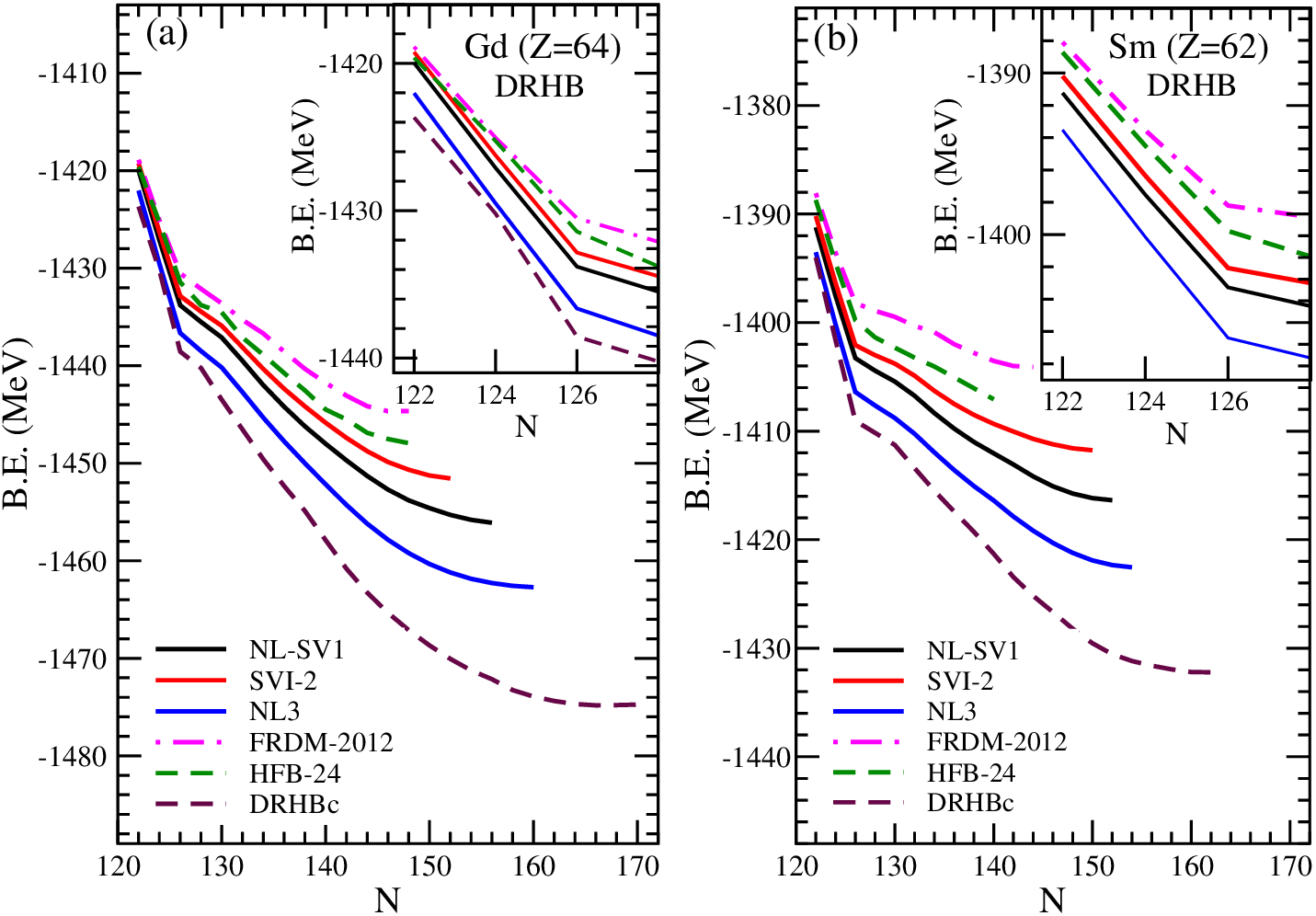}}
\vspace {0.5 cm}
\caption{The same as in Fig.\ref{fig:fig2} for (a) Gd and (b) 
Sm isotopic chains.} 
\label{fig:fig4}       
\end{figure*}
Fig. \ref{fig:fig4} shows the total binding 
energy of nuclei in the Gd ($Z=64$) and Sm ($Z=62$) 
isotopic chains. The divergences between predictions of various models become
stronger as one descends to lower $Z$ values. It is more so for $Z$ values
such as  $Z=62$ (Sm) or even below it for Nd ($Z=60$) and Ce ($Z=58$).
Binding energies for Nd and Ce will not be presented here, for there will be
only a few bound nuclei above $N=126$, and thus the neutron drip line will be
constrained to much lower neutron numbers near $N=126$, as will be seen
in some other figures in this work.

Results with NL-SV1 and SVI-2 show close proximity for Gd isotopes as well as
for Sm isotopes until $N\approx130$. Above this, binding energies 
with NL-SV1 and SVI-2 start diverging, reaching a difference of $\approx 4-5$ MeV.
Here, the results of NL3 show the  divergence of $\approx6$ MeV from NL-SV1
and $\approx 10$ MeV from SVI-2 for the extreme case of drip line
nuclei. In comparison, DRHBc overbinds nuclei excessively also for the
isotopic chains of Gd and Sm, consistent with Fig. \ref{fig:fig2} and
Fig. \ref{fig:fig3}.

A comparison with the mass models (Fig. \ref{fig:fig4}(a)) shows that
at the drip line, 
all the RMF models and mass formula show divergences with each other, which
grow as one traverses from $N\approx140$ to the drip line at $N\approx150$ and above.
The neutron drip line is predicted to fall at 
$N\approx156$ with NL-SV1 and NL3, whereas SVI-2 predicts it to occur at $N\approx154$.
Correspondingly, it falls at $N\approx154$ for HFB-24, whereas FRDM-2012 predicts it
for 4 neutrons less at $N\approx150$. This is consistent with the overall
underbinding of nuclei with FRDM. For Sm isotopes (Fig. \ref{fig:fig4}(b)),
the corresponding drip line is constrained to much lower $N$ number. A common
feature of Figs. \ref{fig:fig2}-\ref{fig:fig4} is that the shell effects in nuclei
at $N=126$ in all the isotopic chains remain strong, as refected by the
kink at $N=126$ (insets of these figures).

\subsection{\label{sec:level3}Quadrupole deformation of nuclei}

\begin{figure}[h]
\resizebox{0.40\textwidth}{!}{%
 {\includegraphics{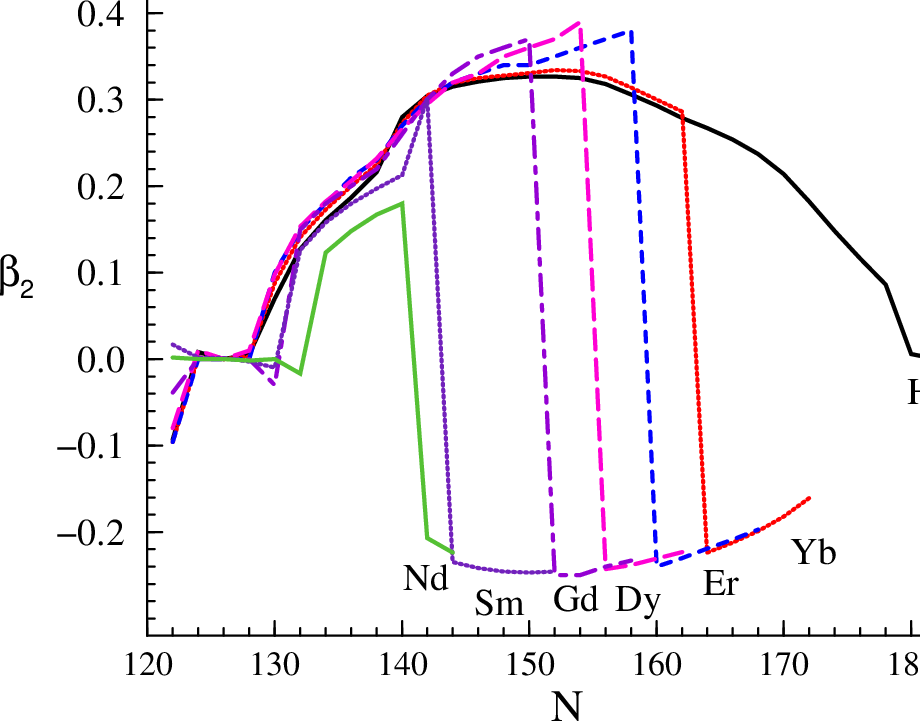}}}
\caption{The quadrupole deformation $\beta_2$ obtained from 
DRHB calculations with the set NL-SV1 of the Lagrangian model 
NL-$\sigma\omega$.}
\label{fig:fig5}
\end{figure}

\begin{figure}[h]
\resizebox{0.45\textwidth}{!}{%
    {\includegraphics{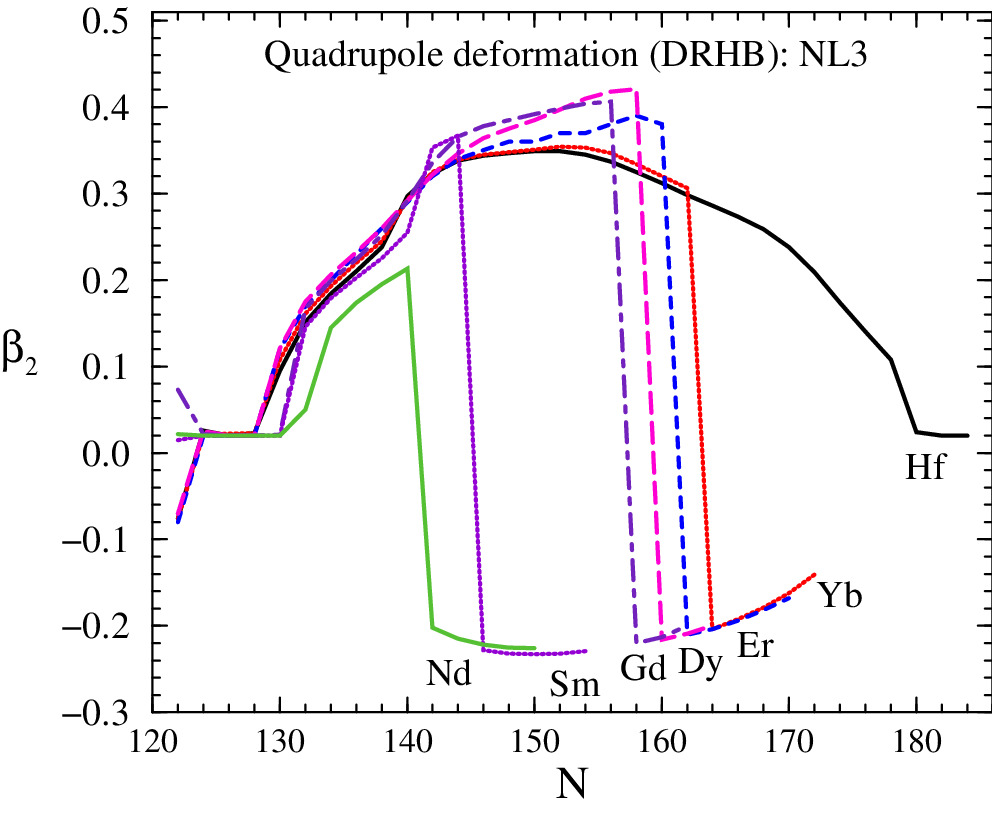}}}
\caption{The same as in Fig. \ref{fig:fig5} but for the set NL3 of the
  Lagrangian model NL-$\sigma$.}
\label{fig:fig6}
\end{figure}

\begin{figure}[h]
\resizebox{0.45\textwidth}{!}{%
     {\includegraphics{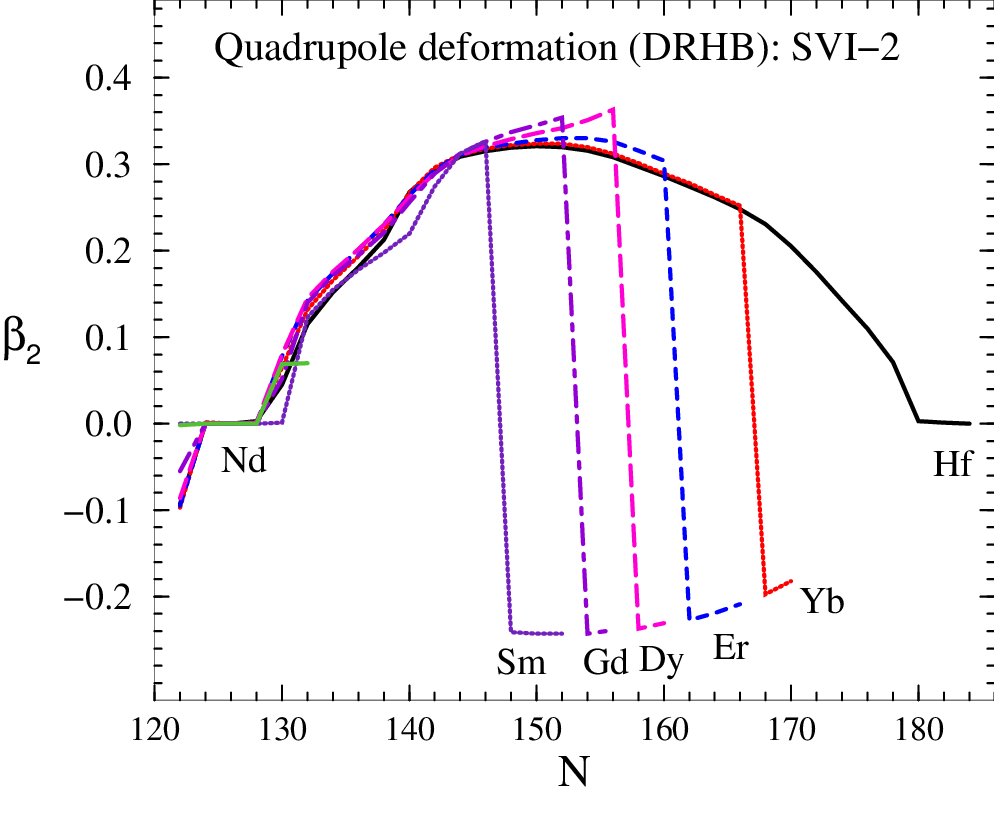}}}
\caption{The same as in Fig. \ref{fig:fig5} but for the set SVI-2 of the
  Lagrangian model SVI.} 
\label{fig:fig7}
\end{figure}

The quadrupole deformation $\beta_2$ of nuclei in various isotopic chains as
obtained from DRHB calculations using various Lagrangian models 
are shown in Figs.~\ref{fig:fig5}-\ref{fig:fig7}. 

Fig.~\ref{fig:fig5} shows $\beta_2$ obtained with NL-SV1.
Beginning with a spherical shape at $N=126$ ($A=186$ for Nd ($Z=60$) 
to $A=192$ for Hf ($Z=72$)), nuclei in various isotopic chains
gain a prolate deformation that increases steadily with an increase in the 
neutron number. The onset of deformation begins at the neutron number $N \approx 134$ 
with a modest value of $\beta_2 \approx 0.12$ for all the isotopic chains.

For the Nd ($Z=60$) chain, $\beta_2$ increases up to 0.18 at $N=138$. 
It makes a shape transition to an oblate shape 
($\beta_2 \approx -0.20)$ at $N=142$, 
before an impending arrival of the drip line at N = 144. For the next chain
of Sm ($Z=62$), nuclei assume an increasigly deformed prolate shape with 
$\beta_2$ going well up to $\approx$0.30 as the neutron number increases. Here, the
maximum in $\beta_2$ is reached at $A=204$ ($N=142$). An abrupt shape transition 
to an oblate shape is seen at $N=144$, two neutrons more than that 
for Nd. The corresponding drip line for Sm is shifted to higher neutron 
number at $N \approx 154$. 

For the higher chains of Gd ($Z=64$), Dy ($Z=66$), Er ($Z=68$) and Yb ($Z=70$),
nuclei assume an increasingly large prolate shape, beginning with an 
onset of deformation at $N=134$. As the $Z$ value of a chain increases, more
and more nuclei exhibit prolate deformation. The maximum in deformation is
reached at $N=150$ for Gd, at $N=154$ for Dy, and at
$N=158$ for Er. The maximum values of $\beta_2$ are: 0.37 for Gd, 0.39 for Dy
and 0.38 for Er. The maximum in deformation in these isotopic chains is
followed immediately by an abrupt shape transition to an oblate shape with
$\beta_2 \approx -0.24$ for almost all the chains except for Hf ($Z=72$). This shape
transition to an oblate shape occurs at an increasingly large neutron numbers
of $N=$152, 156, 160 and 164 for Gd, Dy, Er and Yb isotopic chains,
respectively.

This implies that with a further increase in $Z$ value above Nd ($Z=60$),   
more and more bound nuclei in each chain become available until the maximum
in $\beta_2$ value is attained. The neutron numbers $N=150-158$ are
close to the middle of the shell between $N=126$ and $N=184$. Thus, nuclei
in the chains of Gd ($Z=64$), Dy ($Z=66$), and Er ($Z=68$) are able to reach a 
large maximum in $\beta_2\approx 0.38$ due to proximity to the middle of the shell.
For the next higher chain of Yb ($Z=70$), the maximum in $\beta_2$ is not
sharp, though the maximum in $\beta_2$ itself occurs at $N \approx 150$ closer
to the middle of the shell. However, unlike the other 
chains, the quadrupole deformation tapers to lower values with a shape
transition to an oblate shape at $N=164$. 

The behaviour of Hf ($Z=72$) chain stands out. For this chain, the maximum in
$\beta_2$ does occur near the middle of the shell at $N \approx 154$. The $\beta_2$ 
value decreases gradually with every further increase in the neutron 
number. The deformation decreases eventually to zero as the magic number
$N=184$ approches. Here, one thing that is strikingly different is the 
absence of a shape transition to an oblate shape. Although the whole neutron 
space between $N=126$ and $N=184$ is available for bound nuclei for Hf,
all nuclei assume a prolate shape, giving rise to an inverted parabola. 

The last points in each curve represent nuclei very close to neutron drip line with
$S_{2n} \approx  0$. As seen in Fig. \ref{fig:fig5}, the neutron drip line traverses
through a neutron number that occurs at increasingly higher $N$ values
with every increase in the $Z$ value. The drip line nuclei for all the chains
from Nd ($Z=60$) to Yb ($Z=70$) are oblate with $\beta_2$ in the vicinity of 
$\beta_2 \approx-0.20$. These so-called 'end-point' nuclei are at $N$ = 144, 152, 158, 162, 168,
172 and 184, for Nd, Sm, Gd, Dy, Er, Yb and Hf chains, respectively. Thus, the
neutron drip line traverses from just above $N=126$ for Ce ($Z=58$, not shown 
here) to another magic number $N=184$ for Hf ($Z=72$). For a difference of 14 
protons from $Z=58$ to $Z=72$, the neutron drip line covers a distance of 54
neutrons. Thus, for every increase in the $Z$ values, the domain of bound nuclei
increases enormously. 

The results of deformation obtained with the Lagrangian model NL3 are shown in
Fig. \ref{fig:fig6}. The results with NL3 are very similar to those with the
Lagrangian model NL-SV1 (Fig. \ref{fig:fig5}). NL3 exhibits a maximum in 
$\beta_2$ value followed by an abrupt shape transition to an oblate shape
and a continuation of an oblate shape until the drip line is reached, similar
to that shown by NL-SV1. As noted in Figs. \ref{fig:fig2}-\ref{fig:fig4},
model NL3 was prone to overbinding the nuclei especially in the extreme 
regions, as compared to other Lagrangian models. Consequently, the 
maximum in deformation $\beta_2$ with NL3 is shifted to 
slightly higher neutron numbers. The maximum in deformation with NL3 occurs 
at $N$ = 144, 156, 158, and 160 as against $N$ = 142, 150, 154, and 158, for
Sm ($Z=62$), Gd ($Z=64$), Dy ($Z=66$) and Er ($Z=68$) isotopes, 
respectively. The shift is by 2 neutrons for Sm, 6 neutrons for Gd, 
4 neutrons for Dy and 2 neutrons for Er.

The behaviour of Yb and Hf isotopic chains with NL3 is also very similar
to that with NL-SV1. The maximum is $\beta_2 \approx 0.30$ at $N\approx224$
both for Yb and Hf. The $\beta_2$ tapers off both for Yb
and Hf to lower values, with Yb showing a shape transition to an oblate shape
at N=164, as seen with NL-SV1. Hf chain does not show a shape transition to an oblate
shape. The $\beta_2$ for Hf isotopes above $N=164$ keeps on decreasing
until a spherical shape is reached at $N=184$, as also observed with NL-SV1 in
Fig.~\ref{fig:fig5}.  

The additional binding of nuclei with NL3 has also consequence on 
occurrence of the neutron drip line. The neutron line is correspondingly 
shifted to higher neutron numbers as compared to NL-SV1. The drip line 
with NL3 is shifted by 2-4 neutrons; the shift being higher in the middle 
of the shell. For Yb and Hf chains, the drip line with NL3 coincides with
that with NL-SV1. Thus, even when the total binding energy of Yb and Hf nuclei
near the drip line is higher by $\approx8$ MeV with NL3, the vicinity of the shell
closure $N=184$ to the drip line constrains its extension to higher neutron
numbers than with NL-SV1, especially for the Hf chain. In so far as the shell
effects at $N=184$ are concerned, the drip line for 
Hf isotopes is extended extraordinarily to $N=184$ as compared to that at 
$N=172$ for Yb chain. The shell closure $N=184$ allows a largest number of 
bound nuclei for Hf between the two magic numbers $N=126$ and $N=184$.  

The $\beta_2$ values obtained with the Lagrangian model
SVI-2 are shown in Fig.~\ref{fig:fig7}. All the features shown here
resemble closely to those observed with NL-SV1 (Fig. \ref{fig:fig5}) and NL3
(Fig. \ref{fig:fig6}). It was seen in Figs. \ref{fig:fig2}-\ref{fig:fig4} 
that binding energies with SVI-2 were the lowest amongst all the Lagrangian models in
the region of the neutron drip lines. This has also a consequence that with
SVI-2 a noticeable difference arises for the isotopic chain of Nd ($Z=58$). 
Whilst with NL-SV1 and NL3, deformation keeps rising up to a maximum at 
$N=140$ followed by a shape transition to an oblate shape at $N=142$, Nd nuclei
become unbound already at $N=134$ with SVI-2. Thus, the neutron drip line for Nd chain
has arrived with SVI-2 already at $N=132$, much sooner than with NL-SV1 and NL3.
For the heavier chains of Sm ($Z=62$), Gd ($Z=64$) and Dy ($Z=66$), nuclei gain
deformation as in Figs. \ref{fig:fig5}-\ref{fig:fig6}. However, the 
maximum in $\beta_2$ for these chains with SVI-2 arises at 2 neutrons 
later than that with NL-SV1 and 2 neutrons earlier than with NL3.

\begin{table*}
\caption{\label{tab:table2}The total binding energy (B.E.) and quadrupole
deformation ($\beta_2$) of ground state of nuclei for isotopes of
Er ($Z=68$) obtained with DRHB calculations using model Lagrangians NL-SV1, SVI-2 and
NL3. A comparison is made with the predictions of the mass formulas FRDM-2012,
HFB-24, and DRHBc. The B.E. are in MeV. The standard deviation $\delta$ for
binding energies for a given model with respect to HFB-24 values 
are shown in the bottom line, in units of MeV.}
\begin{ruledtabular}
\begin{tabular}{c c c c r c r c c  c| c r c c lr}
&&&&\multicolumn{2}{c}{RMF Models} &&&&&&\multicolumn{2}{c}{Mass Formulas}&&\\
&  & & NL-SV1 & & SVI-2 & &    NL3 &&&  FRDM-2012 && HFB-24 & &DRHBc& \\
\hline
& $A$ & $N$ & B.E. & $\beta_2$ & B.E. & $\beta_2$ & B.E. 
  &  $\beta_2$ && B.E. & $\beta_2$ & B.E.  &$\beta_2$& B.E.  &$\beta_2$ \\    
\hline 
&194 &126 &-1490.1 &  0.00&-1489.6&0.00 & -1492.7&  0.00&&
                                        -1489.3&0.00&-1489.4&0.00&-1492.9& 0.00\\
&196 &128 &-1493.2 &  0.00&-1492.7&0.00 & -1495.7&  0.00&&
                                        -1492.7&0.00&-1493.5&0.07&-1495.7&0.00\\
&198 &130 &-1496.2 &  0.10&-1495.7&0.08 & -1498.8&  0.10&&
                                        -1495.5&0.07&-1496.7&0.05&-1500.5&0.11 \\
&200 &132 &-1500.2 &  0.15&-1499.5&0.14 & -1502.9&  0.15&&
                                        -1498.9&0.12&-1500.2&0.12&-1505.2&0.16\\
&202 &134 &-1504.2 &  0.18&-1503.3&0.17 & -1506.9&  0.18&&
                                        -1502.0&0.12&-1503.6&0.14&-1509.7&0.19\\
&204 &136 &-1507.9 &  0.21&-1506.8&0.20 & -1510.9&  0.21&&
                                         -1505.6&0.17&-1507.5&0.19&-1513.9&0.21\\
&206 &138 &-1511.4 &  0.23&-1510.2&0.23 & -1514.6&  0.24&&
                                         -1508.8&0.19&-1511.1&0.23&-1517.9&0.23\\
&208 &140 &-1514.7 &  0.27&-1513.4&0.26 & -1518.2&  0.27&&
                                         -1511.8&0.22&-1514.2&0.26&-1521.9&0.27\\
&210 &142 &-1518.0 &  0.30&-1516.5&0.29 & -1521.7&  0.30&&
                                         -1514.7&0.24&-1517.1&0.30&-1526.0&0.38\\
&212 &144 &-1520.9 &  0.32&-1519.3&0.31 & -1524.8&  0.32&&
                                         -1517.2&0.24&-1519.7&0.29&-1529.8&0.33\\
&214 &146 &-1523.5 &  0.33&-1521.7&0.32 & -1527.7&  0.33&&
                                         -1519.3&0.25&-1521.8&0.31&-1533.0&0.34\\
&216 &148 &-1525.7 &  0.34&-1523.8&0.32 & -1530.2&  0.34&&
                                         -1521.1&0.26&-1523.9&0.30&-1536.0&0.34\\
&218 &150 &-1527.7 &  0.34&-1525.5&0.33 & -1532.4&  0.34&&
                                         -1522.5&0.26&-1525.5&0.31&-1538.7&0.34\\
&220 &152 &-1529.3 &  0.35&-1527.0&0.33 & -1534.4&  0.35&&
                                         -1523.4&0.26&-1526.9&0.31&-1541.3&0.34\\
&222 &154 &-1530.7 &  0.36&-1528.2&0.33 & -1536.1&  0.35&&
                                         -1523.7&0.26&-1527.6&0.32&-1543.5&-0.26\\
&224 &156 &-1531.9 &  0.37&-1529.0&0.33 & -1537.5&  0.36&&
                                         -1523.6&0.27&-1528.4&0.31&-1545.8&-0.25\\
&226 &158 &-1532.8 &  0.38&-1529.7&0.32 & -1538.7&  0.37&&
                                         -1523.2&0.27&-1528.6&0.31&-1547.8&-0.25\\
&228 &160 &-1533.7 & -0.24&-1530.1&-0.23 & -1539.7& -0.24&&
                                           &   &-1528.6&0.31 &-1549.6&-0.24  \\
&230 &162 &-1534.3 & -0.23&-1530.6&-0.23 & -1540.7& -0.23&&  &   &    &&-1551.2&-0.24 \\
&232 &164 &-1534.6 & -0.22&-1530.7&-0.22 & -1541.3& -0.22&&  &   &    &&-1552.4&-0.23\\
&234 &166 &-1534.8 & -0.21&         &      & -1541.8& -0.21&& &   &   &&-1553.3&-0.21\\ 
&236 &168 &-1534.7 & -0.20&         &      & -1542.1& -0.20&& &   &    &&-1554.0&0.20 \\
\hline
& $\delta $&& 2.1   &      &    0.7  &      &  6.1   &      &&  2.8   &   & 0.0 &&~~~11.6&\\
\end{tabular}
\end{ruledtabular}
\end{table*}
An abrupt shape transition from a prolate shape to an oblate shape with
SVI-2 for all isotopic chains except for Hf ($Z=72$) is very similar to
that seen with NL-SV1 (Fig.~\ref{fig:fig5}) and NL3 (Fig.~\ref{fig:fig6})
The behaviour of the Hf ($Z=72$) chain with SVI-2 is also similar
to that seen with NL-SV1 and NL3. Hf chain shows no prolate to oblate
transtion in all the models. 

Since nuclei near the neutron drip line with SVI-2 are less bound by $\approx$4-8
MeV as compared to NL-SV1 and NL3, the drip line with SVI-2 is consequently
reached a few neutrons earlier than with NL-SV1 and NL3. Whereas for the 
Sm ($Z=62$) chain, the neutron drip line with SVI-2 coincides with that with
NL-SV1, it reaches two neutrons earlier for Gd, Dy, and Er chains and
four neutrons earlier for Yb chain with SVI-2 as compared to NL-SV1. 
For the Hf chain, the drip line with SVI-2 is reached at the magic number
$N=184$ as is the case with NL-SV1 and NL3. The magic number $N=184$ provides
the boundary of the boundness of Hf isotopic chain.

\subsection{\label{sec:compare}A quantitative comparison of the RMF models with the
  mass formulas}

Three Lagrangian models employed in this work yield results somewhat similar
to each other in a few respects and differ in others. For nuclei that are
relevant to $r$-process nucleosynthesis, i.e. especially those that are
near the $r$-process path, the binding energy, deformation and shape
coexistence are the important properties. Network chain calculations employ
mass formulas which have been refined over a period of a long time to improve
description of known nuclei including those in the extreme regions having
become experimentally accessible. Based upon their success to describe
known nuclei with high precision, FRDM-2012 \cite{Moeller.16}
and HFB-24 \cite{Goriely.13} are two such approaches which make predictions
in the extreme regions upto drip lines. A recent addition to the mass formulas
is DRHBc based upon Deformed Relativistic Hartree-Bogoliub theory including
the continuum effect \cite{Guo.2024}.

FRDM-2012 is a mass formula based upon finite-range droplet model
(macroscopic) superimposed with shell corrections added with Strutinsky 
Integral (a substitute for microscopic shell structure). HFB-24, on the other
hand, is a mass formula based upon microscopic Hartree-Fock Bogoliubov
approach, with 16 parameters fine-tuned over a large number of
attempts ($24^{th}$ in this case). It has thus attained a maturity and
perhaps a level of confidence for a better prediction of the ground state of
nuclei in the extreme regions. 
The microscopic theories have not attained such capabilities as yet.   
 
Here a quantitative comparison of binding energies (B.E.) and 
quadrupole deformation ($\beta_2$) of nuclei in a few isotopic chains
as obtained with various Lagrangian models is made with those with mass formulas
FRDM-2012, HFB-24 and DRHBc. Two representative isotopic chains of
Er ($Z=68$) and Yb ($Z=70$) are taken for comparison. The results for isotopes
of Er with $N \ge 126$ up until the drip line are shown in Table~\ref{tab:table2}.

All the three Lagrangian models predict deformations that agree with 
each other, including on a shape transition from prolate to oblate at $N=160$.
Deformations with RMF models are in general closer to HFB-24, whereas 
FRDM-2012 predicts $\beta_2$ values that are lower by $ \approx 30\%$ than RMF 
models and also are lower than those of HFB-24. The shape transition
in RMF models at $N=160$ is not shown by HFB-24. DRHBc, in comparison,
predicts $\beta_2$ values that are comparable to those of the RMF models.
However, it does predict the shape transition at a lower neutron number
$N=154$ compared to the RMF models.

The binding energies of nuclei with NL-SV1 and SVI-2 are closer to each
other for $N=126-130$ within $\approx 0.5$ MeV. For $N > 130$, NL-SV1 overestimates
the binding energies compared to SVI-2. The overbinding by NL-SV1 keeps 
on increasing as one approaches the neutron drip line, as shown in 
Fig. \ref{fig:fig3}. For instance, NL-SV1 begins overestimating 
SVI-2 values by 1.3 MeV at $N=140$ to 2.2 MeV at $N=150$
and by $3.9$ MeV at $N=164$.

The binding energies from NL3 are compared to those from NL-SV1 and SVI-2 in
Table~\ref{tab:table2}. Based upon very good description of ground-state 
properties along much of the known nuclei, NL3 is considered to be one of 
the best parameter sets within the NL-$\sigma$ model. However, in going to
extreme regions, NL3 predictions deviate strongly from the other models.
NL3 overbinds nuclei significantly as compared to NL-SV1 and
SVI-2. The deviation of NL3 starts at 2.6 MeV with NL-SV1 and 3.1 MeV
with SVI-2 for $N=126$. Overbinding by NL3 as compared to
NL-SV1 and NL3 keeps on increasing in approaching the drip line. 
NL3 overestimates NL-SV1 values by 3.5 MeV at $N=140$, by 4.7 MeV at 
$N=150$ and by 6.0 MeV at $N=160$. A comparison of NL3 values with SVI-2
shows that NL3 overestimates SVI-2 values by 4.8 MeV at $N=140$, by 6.9 
MeV at $N=150$ and by 8.6 MeV at $N=160$. Large differences between NL3 and
other RMF models were seen qualitatively in Fig.~\ref{fig:fig3}. Thus, 
$r$-process nuclei with NL3 are overbound strongly as compared to both NL-SV1 
and SVI-2. Whether overesimation for binding energies of nuclei in the 
$r$-process region by Lagrangian model NL-$\sigma$ is a general 
feature has to be seen.

Comparison of binding energies of mass formulas FRDM-2012 and HFB-24 in 
Table~\ref{tab:table2} shows that they agree for magic nucleus $N=126$. 
However, above $N=126$, FRDM-2012 binding energies are a few MeV less than
those of HFB-24. FRDM-2012 underbinds 
nuclei by 1.2 MeV at $N=130$, by 2.4 MeV at $N=140$, by 3.0 MeV at $N=150$
and by 5.4 MeV at $N=158$ as compared to HFB-24. Thus, extrapolations
in the extreme region are met with large differences. 

A comparison of RMF models with mass formulas (Table~\ref{tab:table2})
shows that NL-SV1 values agree with HFB-24 values within $\approx 0.6$ MeV 
for $N=126-140$. For $N > 140$ NL-SV1 values show an overbinding as compared
to HFB-24. The NL-SV1 values begin diverging from HFB-24, to the extent of 
2.2 MeV for $N=150$ and by 3.1 MeV for $N=160$. On the other hand, a comparison 
of SVI-2 values with HFB-24 shows interesting results. The binding energies 
with SVI-2 show an excellent agreement with HFB-24 within $0.5-1.0$ MeV 
for much of the range. Especially for $N=144-152$, the SVI-2 results
are within $0.1-0.4$ MeV of the HFB-24 values. 

The NL3 binding energies show a strong overbinding of nuclei as compared to 
HFB-24. NL3 overestimates the HFB-24 values by $2.0-3.5$ MeV for $N=126-138$. The 
divergence of NL3 keeps on increasing in moving towards the neutron drip line. 
NL3 overbinds nuclei by 4.0 MeV for $N=140$, by 6.9 MeV for $N=150$, and by 
11.1 MeV for $N=160$. This was duly reflected in the binding energy curve
for NL3 in Fig. \ref{fig:fig3}(a).

\begin{table*}
\caption{\label{tab:table3}The same as in Table~\ref{tab:table2} but for the
isotopic chain of Yb ($Z=70$).}
\begin{ruledtabular}[t]
\begin{tabular}{c c c c r c r c r c | c r c c c r}
&&&&\multicolumn{2}{c}{RMF Models} &&&&&&\multicolumn{2}{l}{Mass Formulas}&&&\\
&  & & NL-SV1 & & SVI-2 & &    NL3 &&&  FRDM-2012 && HFB-24 &&DRHBc \\
\hline
& $A$ & $N$ & B.E. & $\beta_2$ & B.E. & $\beta_2$ & B.E. 
  &  $\beta_2$ && B.E. & $\beta_2$ & B.E.  &$\beta_2$& B.E. & $\beta_2$ \\    
\hline 
&196 &126 &-1516.0 &  0.00& -1515.7 & 0.00 & -1518.4&  0.00&&
                                      -1516.0&0.00&-1515.8&0.00&-1517.8&0.00  \\
  &198 &128 &-1519.8 &  0.00& -1519.6 & 0.00 & -1522.1&  0.00&&
                                      -1520.2&0.01&-1520.7&0.05&-1521.3&0.00  \\
  &200 &130 &-1523.5 &  0.09& -1523.3 & 0.07 & -1525.9&  0.09&&
                                      -1523.7&0.06&-1524.8&0.05&-1526.9&-0.08\\
  &202 &132 &-1528.1 &  0.14& -1527.7 & 0.13 & -1530.5&  0.15&&
                                      -1527.8&0.11&-1528.9&0.10&-1532.00&0.16 \\
  &204 &134 &-1532.7 &  0.17& -1532.1 & 0.17 & -1535.2&  0.18&&
                                      -1531.8&0.12&-1533.0&0.16&-1537.2&0.18\\
  &206 &136 &-1537.1 &  0.20& -1536.3 & 0.19 & -1539.8&  0.20&&
                                      -1535.8&0.16&-1537.7&0.19&-1542.1&0.20 \\
  &208 &138 &-1541.3 &  0.23& -1540.4 & 0.22 & -1544.2&  0.23&&
                                      -1539.9&0.18&-1541.8&0.20&-1546.8&0.22\\
  &210 &140 &-1545.3 &  0.27& -1544.4 & 0.27 & -1548.5&  0.27&&
                                      -1543.7&0.21&-1546.0&0.25&-1551.4&0.26 \\
  &212 &142 &-1549.4 &  0.31& -1548.3 & 0.30 & -1552.7&  0.30&&
                                      -1547.3&0.22&-1549.7&0.25&-1556.2&0.31 \\
  &214 &144 &-1553.0 &  0.32& -1551.8 & 0.31 & -1556.6&  0.32&&
                                      -1550.6&0.24&-1553.0&0.27&-1560.7&0.33 \\
  &216 &146 &-1556.3 &  0.33& -1555.0 & 0.32 & -1560.1&  0.33&&
                                      -1553.5&0.25&-1555.9&0.30&-1564.6&0.33 \\
  &218 &148 &-1559.2 &  0.33& -1557.7 & 0.32 & -1563.3&  0.33&&
                                      -1556.1&0.25&-1558.7&0.30&-1568.1&0.33 \\
  &220 &150 &-1561.7 &  0.33& -1560.2 & 0.32 & -1566.2&  0.33&&
                                      -1558.4&0.25&-1561.2&0.29&-1571.5&0.33\\
  &222 &152 &-1564.0 &  0.33& -1562.2 & 0.32 & -1568.7&  0.33&&
                                      -1560.1&0.26&-1563.4&0.30&-1574.7& 0.33\\
  &224 &154 &-1565.9 &  0.33& -1564.0 & 0.32 & -1571.0&  0.33&&
                                      -1561.2&0.26&-1565.1&0.30&-1577.5&0.33\\
  &226 &156 &-1567.6 &  0.33& -1565.5 & 0.31 & -1573.0&  0.32&&
                                      -1561.9&0.25&-1566.4&0.30&-1580.1&0.32\\
  &228 &158 &-1569.0 &  0.31& -1566.7 & 0.30 & -1574.7&  0.31&&
                                      -1562.4&0.26&-1567.3&0.29&-1582.5&-0.24\\
  &230 &160 &-1570.2 &  0.30& -1567.7 & 0.29 & -1576.2&  0.30&&
                                      -1563.1&0.26&-1568.0&0.29&-1584.9&-0.24\\
  &232 &162 &-1571.2 &  0.29& -1568.5 & 0.28 & -1577.5&  0.29&&
                                      -1563.3&0.24&-1568.2&0.29&-1587.0&-0.23\\
  &234 &164 &-1572.2 & -0.22& -1569.0 & 0.27 & -1578.6& -0.23&&
                                      -1562.5&0.24&-1568.2&0.26&-1588.7&-0.23\\
  &236 &166 &-1572.8 & -0.21& -1569.3 & 0.25 & -1579.5& -0.22&&
              &    &&&-1590.1&-0.22\\
  &238 &168 &-1573.3 & -0.20& -1569.6 &-0.20 & -1580.3& -0.20&&
              &    &&&-1591.3&-0.19\\
  &240 &170 &-1573.5 & -0.18&         &      & -1580.8& -0.19&&
              &    &&&-1590.6&-0.19\\
  &242 &172 &-1573.7 & -0.16&         &      & -1581.3& -0.17&&
              &    &&&-1591.6&-0.18\\
  &244 &174 &-1573.8 &  0.15&         &      & -1581.6&  0.16&&
              &    &&&-1592.4&-0.15\\
  &246 &176 &-1574.0 &  0.12&         &      & -1582.1&  0.12&&
              &    &&&-1593.4&-0.11\\
\hline
& $\delta $&& 1.4 &   &    1.0  &      &  5.1   &  &&  3.3   &   & 0.0 &&~~~10.6&\\
\end{tabular}
\end{ruledtabular}
\end{table*}
It is instructive to compare the RMF models with DRHBc. It is striking that
DRHBc overestimates the binding enegies by large amounts. For instance, DRHBc
overbinds nuclei, by 8.5 MeV for $N=140$, by 13.3 MeV for $N=150$ and by 19.5
MeV for $N=160$ as compared to SVI-2. The overbinding with DRHBc is
lesser as compared to NL-SV1 and NL3. DRHBc overbinds the dripline nucleus
($N=168$) by 12 MeV when compared with NL3, the model that itself overbinds
nuclei signficantly as compared to NL-SV1 and SVI-2.

The difference between various models can be gauged by
$rms$-deviation. Here HFB-24 is chosen as
a reference model. The $rms$-deviation $\delta$~(B.E.) in binding energies 
calculated for each model with respect to HFB-24 values are shown 
in the bottom line of Table~\ref{tab:table2}. It is 2.1 MeV for NL-SV1,
0.7 MeV for SVI-2 and 6.1 MeV for NL3. The $rms$-deviation between the
mass formula FRDM-2012 and HFB-24 is 2.8 MeV. The $rms$-deviation for DRHBc
is inevitably very high at 11.6 MeV as compared to HFB-24. Here, SVI-2 shows 
a close agreement with HFB-24. It may thus have a potential to make
a mass formula using the RMF model without self-couplings of mesons.   

The ground-state B.E. and $\beta_2$ for the isotopes of Yb ($Z=70$) with 
$N > 126$ until the neutron drip line for various models are compared in 
Table~\ref{tab:table3}. Most of the features for this chain are similar 
to those of Er chain shown in Table~\ref{tab:table2}. The binding 
energies of various models, on the other hand, exhibit slightly different 
responses. For Yb isotopes, predictions of NL-SV1 have improved in comparison
to those for the Er chain. The $rms$-deviation of NL-SV1 from HFB-24 for Yb 
isotopes is 1.4 MeV, as compared to 2.1 MeV for the Er isotopes. The SVI-2
energies show a $rms$-deviation of 1.0 MeV from HFB-24 values. This can
be considered to be in close agreement, given that an RMF model is
not a mass formula. NL3, however, overestimates the binding energies with an
$rms$-deviation of 5.1 MeV, which is slightly better than for Er isotopes.
However, the overbinding is as high as 10 MeV for $N=164$ as compared to
HFB-24. DRHBc overbinds the nuclei excessively, as seen earlier, with an 
$rms$-deviation of 10.6 MeV with HFB-24. It overbinds the drip-line ($N=168$)
nucleus by $\approx$20 MeV as compared to SVI-2. 
\begin{figure*}[t]
\resizebox{0.60\textwidth}{!}{%
   \includegraphics{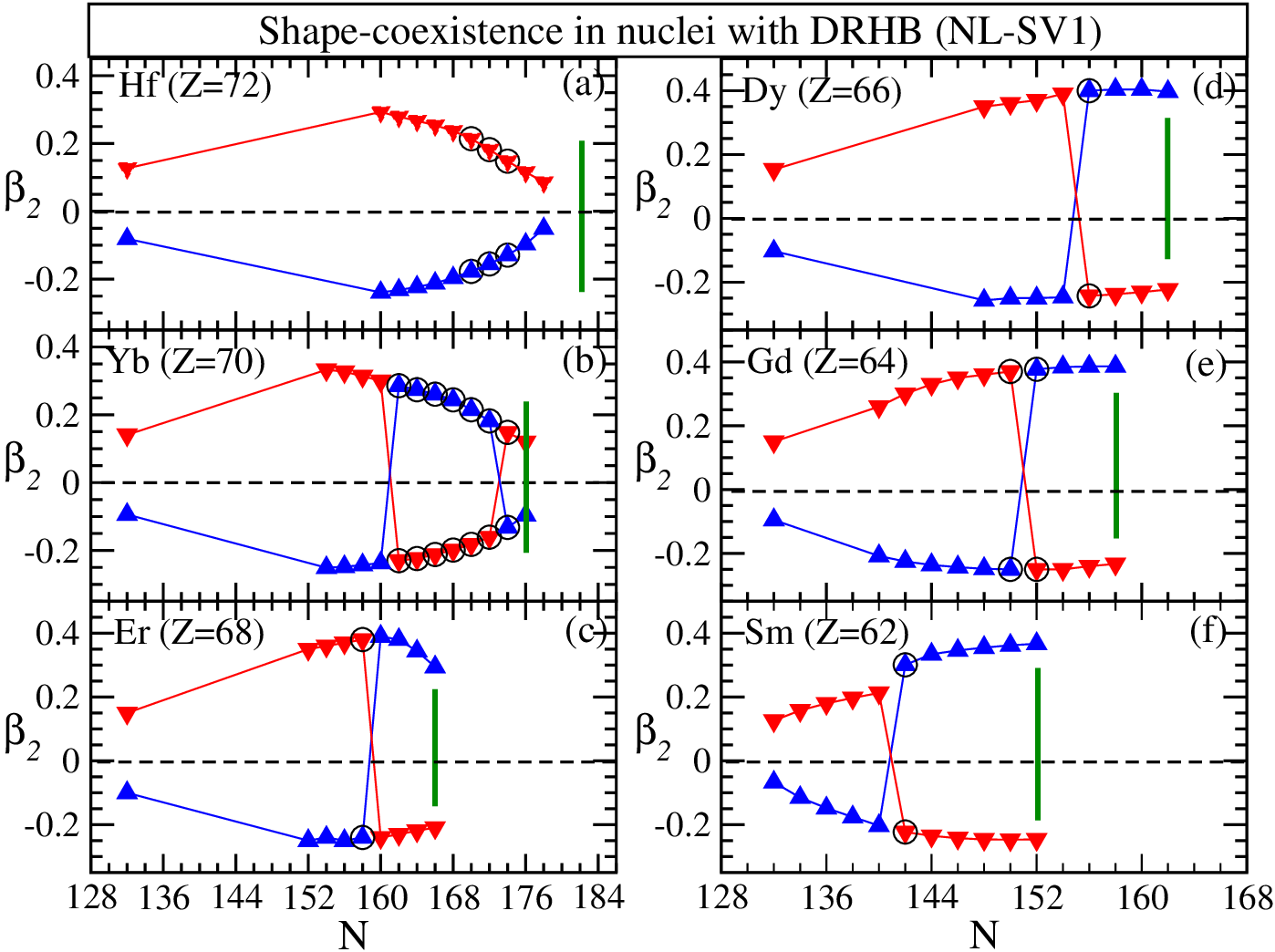}}
\caption{The shape-coexistence of a prolate and an oblate shape in ground
  state of nuclei for the isotopic chains of Hf to Sm as obtained with the
  model NL-SV1. The downward triangle shows the ground state and the upward
  triangle does the second minimum. Nuclei with a ground state
  near-degenerate with a second minimum are encircled.
  The location of the neutron drip
  nuclei is indicated by a vertical bar.} 
\label{fig:fig8}       
\end{figure*}

One feature that is remarkable especially in Table~\ref{tab:table3}   
(seen by the tail of every column) is that both NL-SV1 and NL3 predict 
the neutron drip line extended by 10 neutrons farther than with SVI-2 and 
is 16 neutrons away from the mass formulas. Apparently, both NL-SV1 and NL3
lend additional binding to nuclei so as to stretch the neutron drip line
farther in neutron space closer to the next magic number $N=184$.
In comparison, the neutron drip line with SVI-2 is closest to those with 
FRDM-2012 and HFB-24, and is only 4 neutrons away from that of mass formulas.

The binding energies predicted by FRDM-2012 (Table~\ref{tab:table3})
are $\approx$0.5-1.0 MeV smaller than
those of HFB-24 but agree with those from SVI-2 in the region $N=128-134$.
Above $N=134$, FRDM values underpredict the binding energies both from
SVI-2 and HFB-24. The deficit of B.E. with FDM-2012 is ~5.0 MeV in the
neighbourhood of the neutron drip line as compared to both HFB-24 and SVI-2. 
This difference in binding energies near the drip line can have a substantial 
difference in predictions of $r$-process nuclear abundances. 

\subsection{\label{sec:level4}The shape-coexistence, the 
neutron drip-line and degeneracy}

Interplay between single-particle levels and deformation is known to
produce coexistence between a prolate and an oblate shape in a large
number of nuclei across the periodic table. Occasionally a shape-coexistence
may be punctuated by a degenerate or a near-degenerate second minimum. Here,
we present the results on shape-coexistence, on
near-degeneracy of prolate and oblate shapes in the ground state and occurrence
of such features near the neutron drip-line with out three Lagrangian
models.

In Fig.~\ref{fig:fig8}, nuclei with a ground-state in shape-coexistence
with a second minimum observed with the Lagrangian model NL-SV1 for the
isotopic chains from Hf ($Z=72$) to Sm ($Z=62$) are shown.
As the convention goes, only those nuclei are shown wherein a second minimum 
lies within $\approx$1 MeV of the ground state. The downward triangles show the
$\beta_2$ value for the ground state (the lowest minimum) of nuclei.
This mirrors the quadrupole deformation shown in Fig.~\ref{fig:fig5}. 
The shape transition from a prolate to an oblate shape is clearly
evident in the neutron-rich region for all chains except in Hf.

The upward triangles show the deformation for the second minimum in the 
total binding energy. The shape-coexistence is seen for a large number of 
nuclei in all the isotopic chains. It is noteworthy that the region of neutrons 
$N=132-158$ is devoid of the shape-coexistence for the Hf chain. Here, the 
onset of shape-coexistence starts at $N=162$. For Yb ($Z=70$), Er ($Z=68$), and 
Dy ($Z=66$), the onset of the shape-coexistence begins at $N$ = 154, 152 and 148,
respectively. For the Gd chain, the region of shape-coexistence 
has become much narrower, and the shape-coexistence begins much earlier
at $N=140$. The case of Sm chain is unique. Here, all the nuclei exhibit the
phenomenon of shape-ceoxistence, beginning already at $N=130$.

This region of nuclei is prone to occurrence of shape-coexistence as 
seen in all the figures of Fig.~\ref{fig:fig8}. 
Eleven isotopes of Hf ($Z=72$), thirteen isotopes of Yb ($Z=70$), 
nine isotopes of Er ($Z=68$), nine isotopes of Dy ($Z=66$), 
eleven isotopes of Gd ($Z=64$) and eleven isotopes of Sm ($Z=62$) exhibit
the phenomenon of shape-coexistence. Thus, there is a prepronderance 
of shape-coxistence of prolate-oblate shape in this region of the periodic 
table.

Since deformed RHB calculations presented in this work were carried out until
the neutron drip line is reached for each isotopic chain, it gives a
perspective as to how nuclear properties especially 
deformations evolve near the $r$-process path and near 
the drip-line. Drip line nuclei with $S_{2n} \approx 0$ are 
represented by a vertical line in  Fig.~\ref{fig:fig8}. 
The drip line for Hf is at $N=182$, that is just two neutrons 
below the shell closure N=184, obviating the shape
to be spherical. However, as one goes down in the $Z$ number, the drip 
line starts drifting downwards from $N=184$ and correspondingly nuclei at the
neutron drip line acquire a deformed shape. For the Yb chain, the drip line
nucleus ($N=176$) has a modest prolate deformation of $\beta_2 \approx 0.13$ in
coexistence with an oblate shape. 
\begin{figure}[t]
\resizebox{0.40\textwidth}{!}{%
   \includegraphics{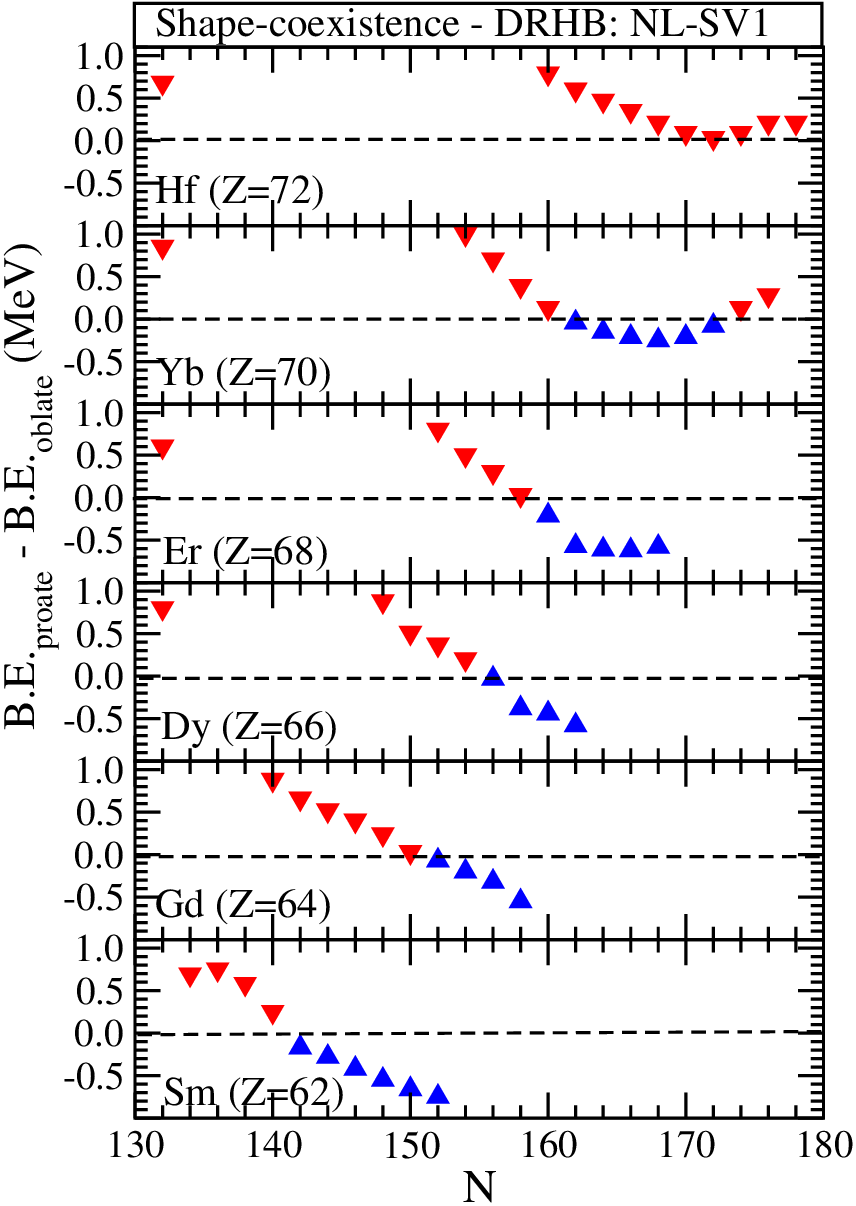}}
\caption{The difference in B.E. of shape-coexistent prolate and oblate
shape in ground state of nuclei for the isotopic chains of Hf to Sm 
obtained with model NL-SV1. The downward triangles (red) show a prolate 
ground state and the upward triagles (blue) show an oblate ground state.} 
\label{fig:fig9}       
\end{figure} 

The drip line occurs at $N=$ 176, 166, 162, 158, and 152 for Yb ($Z=70$), 
Er ($Z=68$), Dy ($Z=66$), Gd ($Z=64$), and Sm ($Z=62$) chains, respectively.
For all these chains, there is a shape-coxistence between a 
prolate and oblate shape at the drip line. Moreover, the ground 
state of the drip line nuclei for all the chains from $Z=62-68$ 
is oblate. The deformation of the oblate ground state of these 
nuclei stays in the vicinity of $\beta_2 \approx 0.24$ in the 
Er ($Z=68$), Dy ($Z=66$), Gd ($Z=64$) and Sm ($Z=62$) chains.    

A significant number of nuclei exhibiting the shape-existence have the second
minimum lying within $\approx$0.1-0.2 MeV of the ground state (absolute minimum). 
Such states are near-degenerate with the ground state. These are shown 
by a cirle encompassing the data points in Fig.~\ref{fig:fig8}. 
The degeneracy is seen in three nuclei with $N=$ 170, 172 and 174 in the 
Hf chain in Fig.~\ref{fig:fig8}(a). The degeneracy is seen
most prominently in the Yb chain Fig.~\ref{fig:fig8}(b), where seven
nuclei with $N=162-174$ exhibit it. The degeneracy begins
at $N=162$ where ground state takes the oblate shape after the shape
transition. Interestingly, a further reversal of the ground-state
shape occurs at $N=174$ in Yb which is prolate shaped. These
nuclei are very close to the $r$-process path. For the isotopic 
chains of Er ($Z=68$), Dy ($Z=66$), Gd ($Z=64$) and Sm ($Z=52$), either one or 
two isotopes exhibit the degeneracy. 

The degree of shape-coexistence in the ground-state of nuclei
is shown in Fig.~\ref{fig:fig9}, where the difference in the 
binding energies of nuclei in the prolate and oblate shapes
$\Delta E =$ B.E. (prolate) - B.E. (oblate) is shown. Positive values shown by the
downward triangles (red in the colored figures) indicate that an oblate shape
is $\Delta E$ above a prolate ground state. The negative values shown by the
upward triangles (blue in the colored figures) denote a prolate state is 
$\Delta E$ above an oblate ground state.

\begin{figure*}[ht]
\resizebox{0.60\textwidth}{!}{%
   \includegraphics{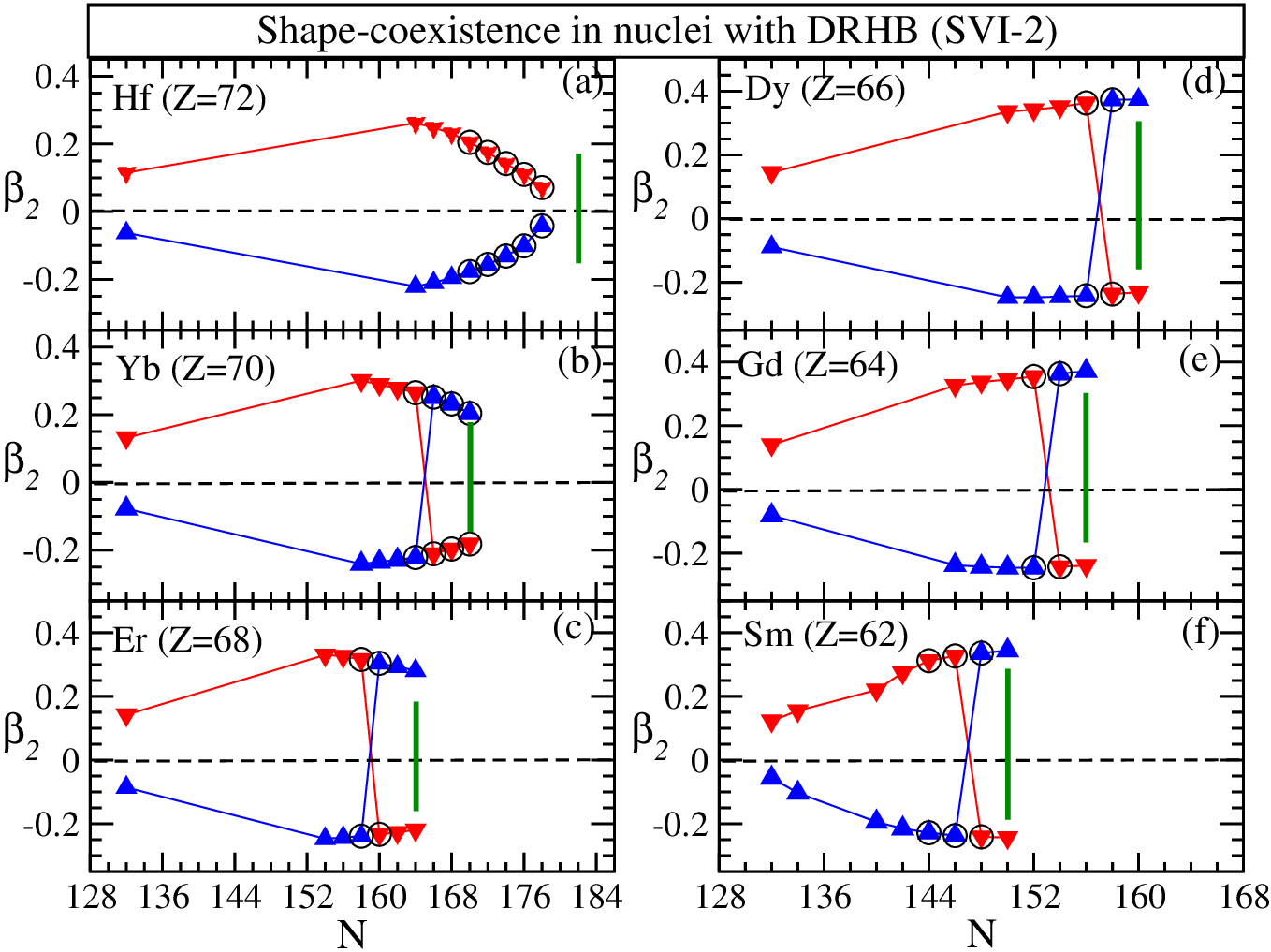}}
\caption{The same as in Fig.~\ref{fig:fig8} but for the Lagrangian model 
SVI-2.} 
\label{fig:fig10}       
\end{figure*}

\begin{table*}
\caption{\label{tab:table4} Shape-coexistence in Hf ($Z=72$) isotopes with the 
RMF models NL-SV1 and SVI-2. The total binding energy (B.E.) in MeV and 
quadrupole deformation $\beta_2$ for the ground state (G.S.) and the 
shape-existing state second minimum (S.M.) as obtained with DRHB are
shown. The energy difference $\Delta{E}$ = B.E. (prolate) - B.E. (oblate) (in MeV)
between the G.S. and S.M. are also shown.}
\begin{ruledtabular}
\begin{tabular}{c c c c c c c c c c c c c c c}
&  & & &&  NL-SV1 &&&&&& SVI-2  && \\
\hline
&  & & G.S. &&  S.M. && &&G.S. & & S.M. && &\\
\hline
& $A$ & $N$ & B.E. & $\beta_2$ & B.E. & $\beta_2$ &$\Delta{E}$ && B.E. 
  &  $\beta_2$ & B.E. & $\beta_2$ &$\Delta{E}$& \\    
\hline 
&232 &160 &-1605.7 & 0.29 & -1604.9& -0.24 &0.79 &&&&&\\
&234 &162 &-1607.3 & 0.28 & -1606.7& -0.23 &0.60 &&&&&\\
&236 &164 &-1608.6 & 0.27 & -1608.1& -0.22 &0.47 && -1606.2 &0.26 &-1605.4 &-0.22&0.74&\\
&238 &166 &-1609.6 & 0.25 & -1609.3& -0.21 &0.35 && -1607.0 &0.25 &-1606.4 &-0.21&0.54&\\
&240 &168 &-1610.5 & 0.24 & -1610.3& -0.20 &0.21 && -1607.6 &0.23 &-1607.2 &-0.20&0.32&\\
&242 &170 &-1611.2 & 0.21 & -1611.1& -0.18 &0.09 && -1608.0 &0.21 &-1607.9 &-0.18&0.10&\\
&244 &172 &-1611.8 & 0.18 & -1611.7& -0.16 &0.03 && -1608.3 &0.18 &-1608.4 &-0.16&0.0&\\
&246 &174 &-1612.4 & 0.15 & -1612.3& -0.13 &0.09 && -1608.7 &0.14 &-1608.7 &-0.13&0.0&\\
&248 &176 &-1613.0 & 0.12 & -1612.8& -0.10 &0.21 && -1609.1 &0.11 &-1609.1 &-0.10&0.06&\\
&250 &178 &-1613.6 & 0.09 & -1613.4& -0.05 &0.21 && -1609.6 &0.07 &-1609.5 &-0.04&0.07\\
\end{tabular}
\end{ruledtabular}
\end{table*}

This figure shows some interesting phenomenon. Leaving aside the point  
at $N=132$ in chains, the shape-coexistence begins at larger $N$ for most of the 
heavier chains as that seen in Fig.~\ref{fig:fig8}. For all the isotopic
chains the region of shape-coexistance begins with a prolate ground-state
where an oblate state is $\Delta E\approx 0.8-1.0$ MeV above the ground state.
The degree of the shape-coexistence is enhanced as further neutrons are
added (indicated by a downward curve) until a region close to 
$\Delta E \approx -0.20-0.20$ MeV is reached, that can be termed as degeneracy or
a near degeneracy of prolate and oblate shapes. This is followed by
the ground state turning to become oblate (with the exception of Hf chain)
with a prolate state above it. Only in case of Yb, there is a further
shape change to a prolate one in the ground state passing through 
a degeneracy in the ground state at a certain neutron number.
For all the other chains, the ground state
becomes an oblate shape (upward triangles) for the higher
neutron numbers with the second minimum (prolate) at increasingly
higher in energy uo to ~0.5-0.7 MeV. An oblate ground-state in nuclei 
near the $r$-process path can have far reaching consequences on 
$\beta$-decay half lives in the $r$-process nucleosynthesis.

The sequence of a shape transition from a prolate ground state 
to an oblate ground state passing through a near degeneracy 
between the two as predicted by the Lagrangian model NL-SV1 
(represented by a nearly srainght line in Fig.~\ref{fig:fig9} 
except for Hf) is an interesting feature of the rare-earth nuclei in the
extreme region of the $r$-process path. This may imply that the 
single-particle structure in the shell between $N=126$ and $N=184$ is 
malleable to smooth changes in the shape of the ground state.

\begin{table*}
  \caption{\label{tab:table5} The same as in Table \ref{tab:table4} but
 for the isotopic chain of Yb ($Z=70$).}
\begin{ruledtabular}
\begin{tabular}{c c c c c c c c c c c c c c c}
&  & & &&  NL-SV1 &&&&&& SVI-2 & & &  \\
\hline
&  & & G.S. &&  S.M. && && G.S. & & S.M. & & &\\
\hline
& $A$ & $N$ & B.E. & $\beta_2$ & B.E. & $\beta_2$& $\Delta{E}$&& B.E. 
  &  $\beta_2$ & B.E. & $\beta_2$ & $\Delta{E}$&\\    
\hline 
&224 &154 &-1565.9 & 0.33 & -1564.9& -0.25 &1.00 & &&&&&&\\
&226 &156 &-1567.6 & 0.33 & -1566.9& -0.25 &0.70 & &&&&&&\\
&228 &158 &-1569.0 & 0.31 & -1568.6& -0.24 &0.39 & & -1566.7 &0.30 &-1565.9 &-0.24&0.81&\\
&230 &160 &-1570.2 & 0.30 & -1570.1& -0.24 &0.13 & & -1567.7 &0.29 &-1567.2 &-0.24&0.54&\\
&232 &162 &-1571.2 & 0.29 & -1571.2& -0.23 &-0.04& & -1568.5 &0.28 &-1568.2 &-0.23&0.30&\\
&234 &164 &-1572.2 &-0.22 & -1572.0&  0.27 &-0.15& & -1569.0 &0.27 &-1568.9 &-0.22&0.14&\\
&236 &166 &-1572.8 &-0.21 & -1572.6&  0.26 &-0.21& & -1569.4 &-0.21&-1569.3 & 0.25&-0.01&\\
&238 &168 &-1573.3 &-0.20 & -1573.0&  0.24 &-0.25& & -1569.6 &-0.20&-1569.5 & 0.23&-0.14&\\
&240 &170 &-1573.5 &-0.18 & -1573.3&  0.22 &-0.21& & -1569.7 &-0.18&-1569.5 & 0.20&-0.21&\\
&242 &172 &-1573.7 &-0.16 & -1573.6&  0.18 &-0.08& &&&&&&\\
&244 &174 &-1573.8 & 0.15 & -1573.7& -0.13 &0.13 & &&&&&&\\
&246 &176 &-1574.0 & 0.12 & -1573.7& -0.10 &0.28 & &&&&&&\\
\end{tabular}
\end{ruledtabular}
\end{table*}

The results obtained on the shape-coexistence within the Lagrangian model SVI-2
are shown in Fig.~\ref{fig:fig10}. All features seen with NL-SV1 
are also seen with SVI-2, with some differences observed at a few places. 
The phenomenon of shape-coexistence between a prolate and an 
oblate shape is seen abundantly also with SVI-2 as was the
case with NL-SV1. However, as compared to NL-SV1, there are
a fewer number of cases of shape-coexistence for each chain. For 
instance, there is bigger void in shape-coexistence between $N=130$ and 
$N=164$ for the Hf chain (Fig.~\ref{fig:fig10}(a)) as compared to that 
with NL-SV1 (Fig.~\ref{fig:fig8}(a)). The region of shape 
coexistence with SVI-2 begins at $N=164$ for the Hf chain as 
compared to that with NL-SV1, where it begins at $N=160$. 
SVI-2 thus exhibits the shape-coexistence in 8 nuclei of Hf 
as compared to that in 10 nuclei with NL-SV1.

The most striking difference is seen for the Yb chain 
(Fig.~\ref{fig:fig10}(b)), where there are only 8 cases of 
shape-coexistence as compared to 13 cases with NL-SV1. For the Er and Dy
chains, SVI-2 shows the shape-coexistence in 2 lesser number of nuclei as
compared to that with NL-SV1. Another major difference is noted for the Gd
chain (Fig.~\ref{fig:fig10}(e)). SVI-2 shows the shape-coexistence in 7
nuclei of Gd as compared to in 11 nuclei with NL-SV1. This may imply that 
the shell structure in the middle of the shell between $N=126$ and $N=184$ 
is not so much soft with SVI-2 as with NL-SV1. 

\begin{figure}
\resizebox{0.40\textwidth}{!}{%
   \includegraphics{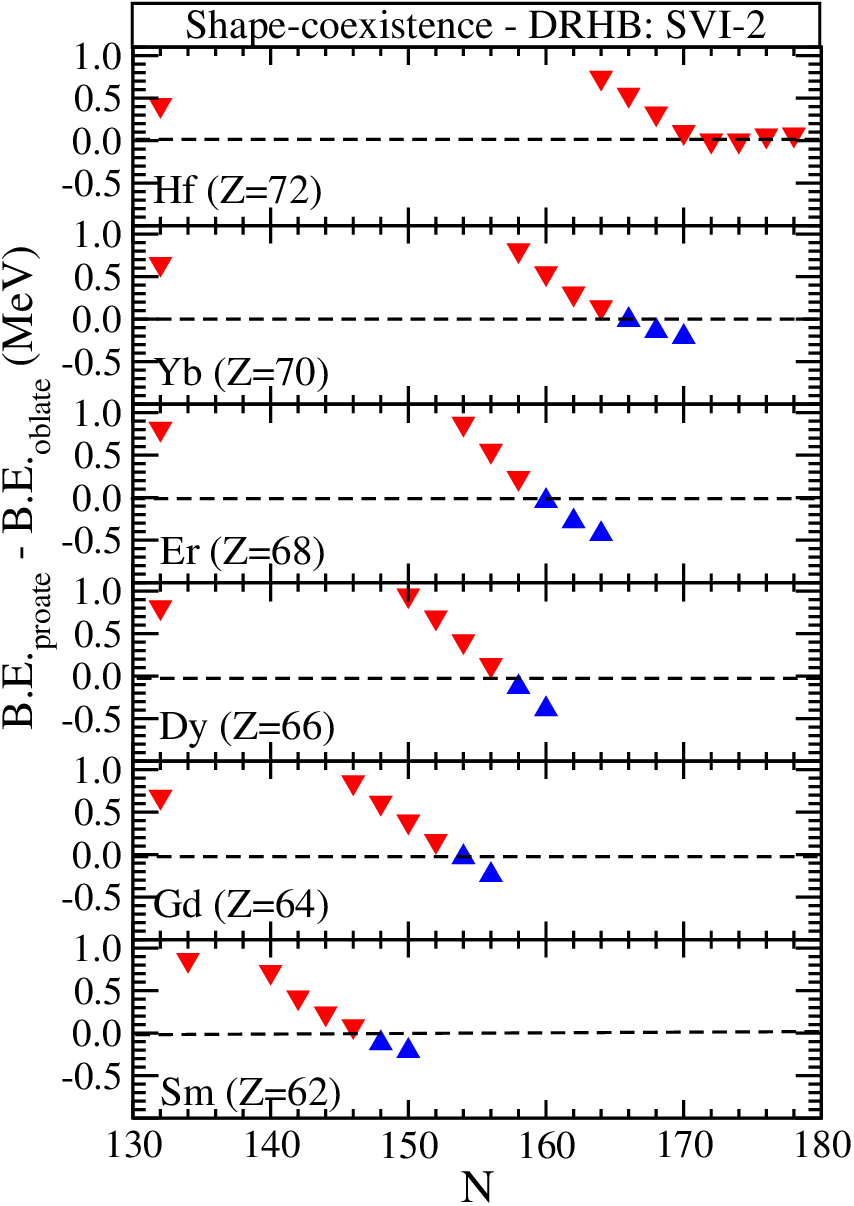}}
\caption{The same as in Fig.~\ref{fig:fig9} but for SVI-2.} 
\label{fig:fig11}       
\end{figure}
The shape-coexistence between a prolate and oblate shape as obtained
with the Lagrangian model SVI-2 is illustrated in Fig.~\ref{fig:fig11}.
The shape-coexistence makes a smooth transition from a prolate ground
state to an oblate ground state for all the chains except for the Hf chain,
as also seen with NL-SV1 in Fig.~\ref{fig:fig9}, but for a fewer isotopes.
The double transformation from prolate to oblate to prolate ground state
for Yb as seen with NL-SV1 is not produced by SVI-2. For all other chains
(except Hf) a smooth walk-over from a prolate to an oblate shape 
passing through a nearly degenerate shapes is clearly demonstrated 
by SVI-2. 

The shape-coexistance in Hf ($Z=72$) and Yb ($Z=70$) chains with NL-SV1 and
SVI-2 are quatitively summarized in Tables ~\ref{tab:table4} and
\ref{tab:table5}, respectively. There are many nuclei that exhibit near
degeneracy in the ground state.

\begin{figure*}
\resizebox{0.60\textwidth}{!}{%
   \includegraphics{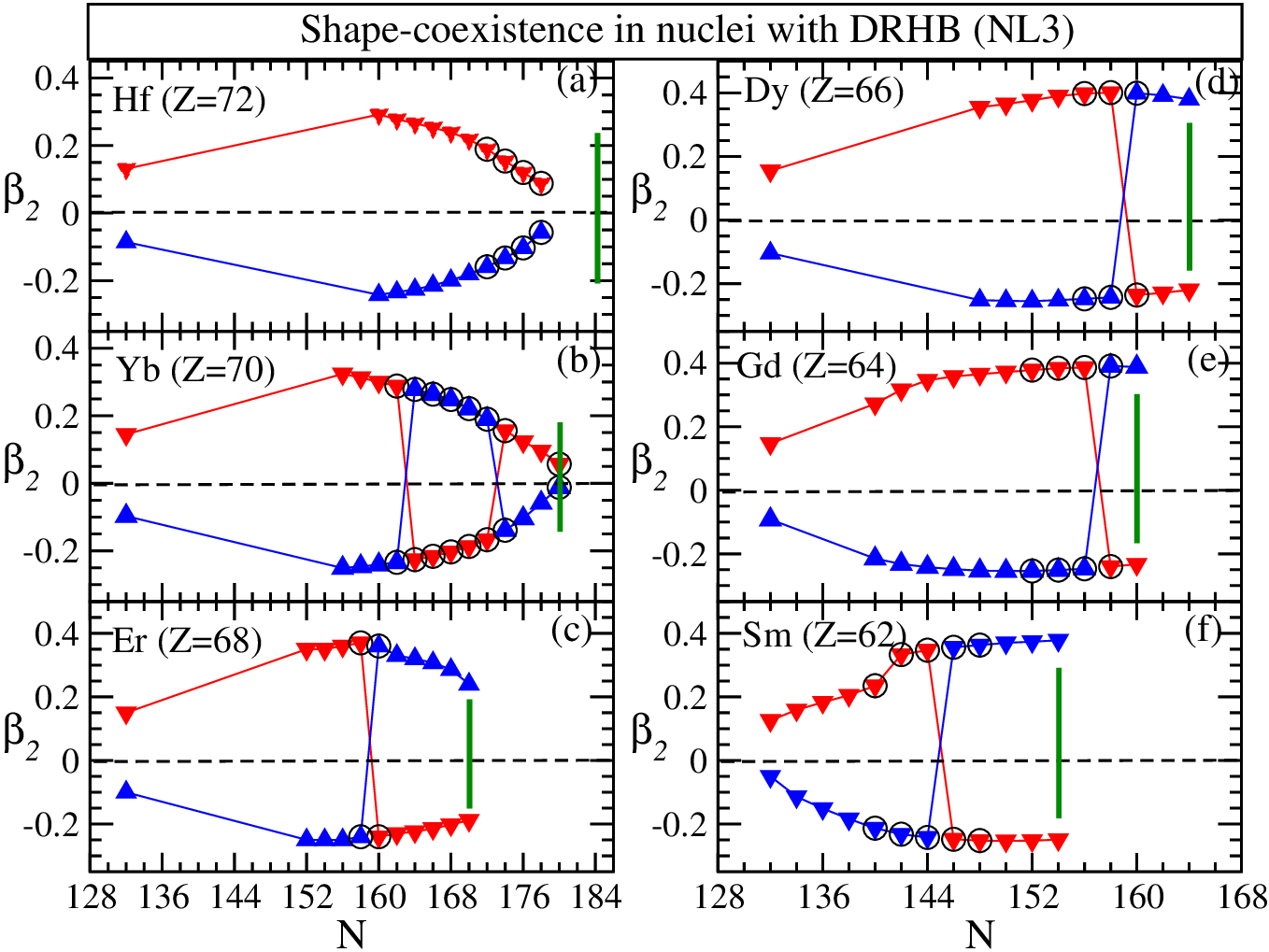}}
\caption{The same as in Fig.~\ref{fig:fig8} but for the Lagrangian model NL3.} 
\label{fig:fig12}       
\end{figure*}

The shape-coexistence along with the corresponding $\beta_2$ values
obtained with the Lagrangian model NL3 
are shown in Fig.~\ref{fig:fig12}. All the features that were
illustrated in Figs.~\ref{fig:fig8} and \ref{fig:fig10} 
for NL-SV1 and SVI-2, respectively, are present with NL3. 
The phenomenon of shape-coexistence between prolate and oblate shape, 
shape transition from prolate to oblate shape, and the oblate 
ground state for Er, Dy, Gd and Sm isotopes for neutron drip line 
nuclei are all reproduced with NL3 as were with 
NL-SV1 and SVI-2. Since NL3 overbinds nuclei near the drip line 
as compared to NL-SV1 and especially as compared to SVI-2, the neutron drip 
line as indicated by vertical bars is shifted up by a few 
neutron mumbers.

The drip line nucleus for Hf ($Z=72$) with NL3 is 
now at the magic number $N=184$ as compared to $N=182$ for NL-SV1 and SVI-2. 
This is a curious case where the neutron drip line has reached a 
magic number. The corresponding drip-line nucleus for Yb with 
NL3 is now nearly spherical as the drip line is shifted by 4 neutrons
up to $N=180$ as against to $N=176$ with NL-SV1 and
to $N=170$ with SVI-2. The neutron drip line for Er, Dy, Gd and Sm chains 
with NL3 is shifted up by 2-4 neutrons as compared to NL-SV1.
The number of shape-coexisting nuclei with NL3 is higher
than with NL-SV1 and much higher than with SVI-2. A larger occurrence of 
shape-coexisting nuclei with NL3 may imply that the shell structure 
with NL3 is softer than that with SVI-2.

The shape-coexistence of prolate and oblate nuclei is expected to influence
abundances in $r$-process nucleosynthesis. The $\beta$-decay half-lives are
sensitive to the shape and deformation of nuclei. The half-lives decrease
significantly as deformation increases. Moreover, a prolate nucleus is
predicted to have a shorter half-life as compared to an oblate nucleus. 
Nuclei with shorter half-lives would contribute more to abundances. For nuclei
with a shape-coexistence, the total half-lives would be a combination of the
two shapes.

\subsection{\label{sec:s2n}Two-neutron separation 
energies and shell effects in nuclei}

Appearance of a kink or it's disappearance at a magic number is reflective
of robustness or otherwise of a shell closure. Using the total binding energies
of nuclei obtained with DRHB shown in Figs.~\ref{fig:fig2}-\ref{fig:fig4}, 
the two-neutron separation energies $S_{2n}$ have been calculated as given by

\begin{equation}
\label{s2n}
S_{2n}~(N, Z) = {\rm{B.E.}}~(N, Z) - {\rm{B.E.}}~(N-2, Z).
\end{equation}

\begin{figure}[b]
\hspace{-1 cm}
\resizebox{0.40\textwidth}{!}{%
   \includegraphics{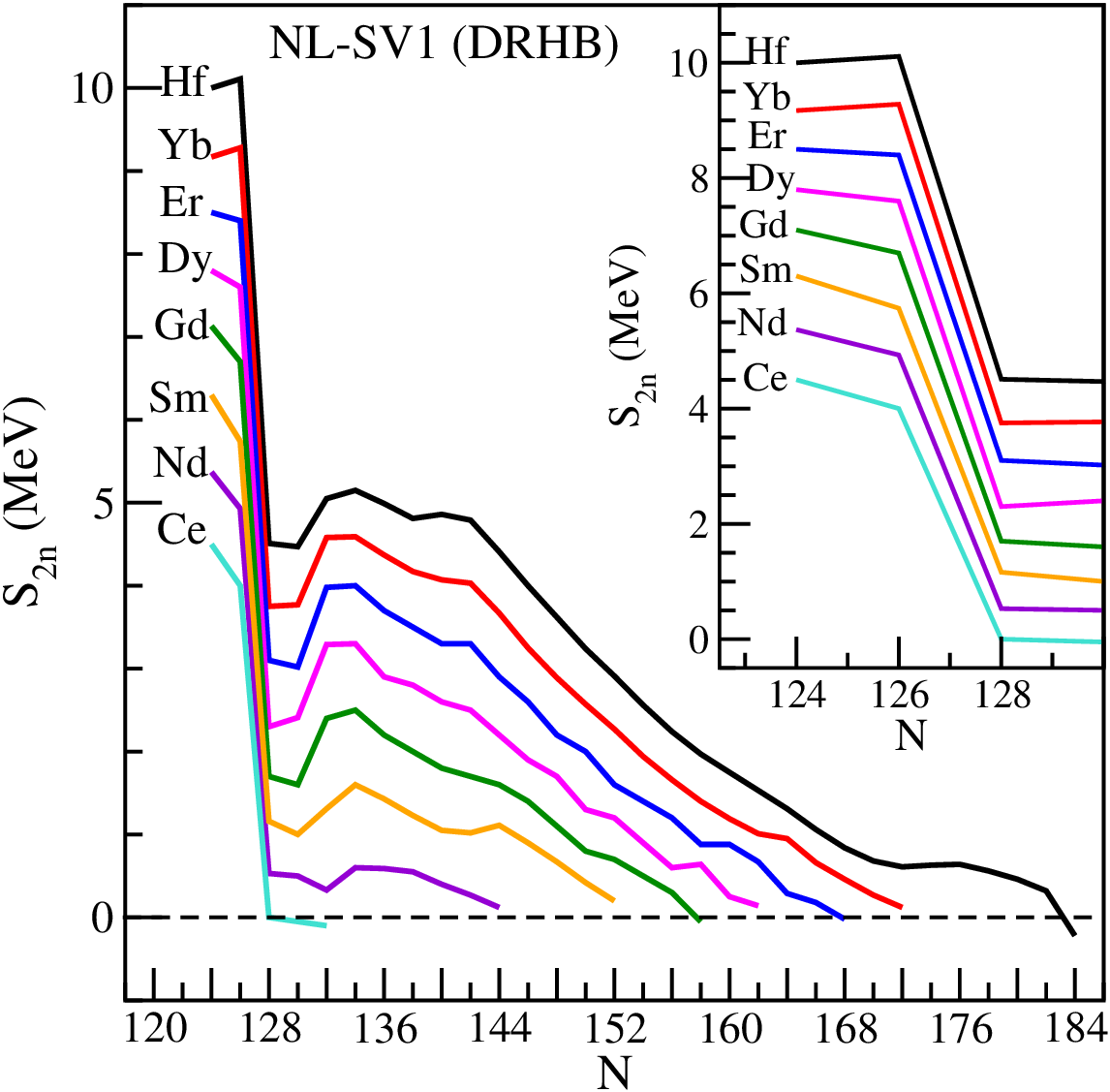}}
\caption{The 2-neutron separation energy $S_{2n}$ of nuclei in the 
isotopic chains $Z=58-68$ obtained from DRHB calculations
using the set NL-SV1 of Lagrangian model NL-$\sigma\omega$.}
\label{fig:fig13}       
\end{figure}

The $S_{2n}$ values obtained with the Lagrangian model NL-SV1 are shown
in Fig.~\ref{fig:fig13}. The strong kink at the magic number $N=126$ is 
seen conspicuosly for all isotopic chains. An abrupt decline in $S_{2n}$ values
at $N=128$ is a manifestation of the strong shell closure at $N=126$.
There is hardly any change in the slope of the curves at $N=126$ 
from Hf ($Z=72$) to Ce ($Z=58$). This is clearly evident
in the inset of the figure, where $S_{2n}$ values in the vicinity
of $N=126$ are shown. 

The $S_{2n}$ value at $N=128$ in Fig.~\ref{fig:fig13}
shows a constant decline from ~4.5 MeV for
Hf ($Z=72$) to nearly zero for Ce ($Z=58$), signaling the 
arrival of the neutron drip line. There are, therefore, no bound Ce 
isotopes beyond N=128. A zero value of 
$S_{2n}$ for Ce is preceded by a smaller value for Nd ($Z=60$) and Sm ($Z=62$). 
For the Nd chain, the $S_{2n}$ value at $N=128$ is $\approx$ 0.5 MeV that decreases 
further to $\approx $0.40 MeV at $N=130$. It is expected to decrease 
further to zero with further addition of neutrons. However, an 
onset of deformation at $N=132$ lends an extra binding energy to 
Nd nuclei at $N=$ 134-140, thus creating a small ridge in the 
$S_{2n}$ curve. In place of an expected neutron drip line in the vicinity
of $N \approx$ 134-136, a ridge of stability is created by the deformation 
in nuclei. However, the additional binding due to deformation can not 
compete for too long and the drip line for Nd is arrived faster at $N=144$, 
yet extending it by about 8-10 neutrons from an expected place.

For the Sm chain, the $S_{2n}$ value at $N=128$ is $\approx 2$ MeV.
With further addition of neutrons it should have decreased significantly.
However, as an onset of deformation takes place at $N=132$ 
(as seen in Fig.~\ref{fig:fig5}), the  $S_{2n}$ values 
begin to increase as for Nd chain. A further 
addition of neutrons creates Sm isotopes with increasing deformation
that adds extra binding energy to nuclei. Interplay between deformation 
energy and symmetry energy manages to create a longer ridge that extends 
from $N=130$ until the neutron drip line is extended to a much higher neutron
number $N=152$ as compared to Nd. The number of bound nuclei with Sm ($Z=58$) is 
much larger than that with Nd ($Z=56$).

The first peak in the ridge in the Sm chain at $N=134$ corresponds 
to an increase in the binding energy by a sudden onset of
deformation at $N=132$ that increases at $N=134$. However, with a further addition
of neutrons, the increase in the corresponding binding energy decreases 
even when the deformation increases and shows a maximum at $N=144$. A shape
transition from a prolate shape to an oblate shape at $N=144$ results in a small
second peak at $N=144$.

For the heavier isotopic chains of Gd ($Z=64$), Dy ($Z=66$), Er ($Z=58$),
Yb ($Z=70$) and Hf ($Z=72$), the process of creation of a ridge begins at
$N=132$ with an onset of deformation, as was seen with Nd ($Z=60$) and Sm ($Z=62$)
above. Instead of an expected neutron drip line above $N=130$, where $S_{2n}$ values should
have fallen down steadily to zero, the deformation creates the available 
isospin space for bound nuclei, which keeps growing with every increase
in the $Z$ value. The growing number of bound nuclei in each successive chain
is engendered by an increasing deformation as shown in Fig.~\ref{fig:fig5}.
Deformation imparts an extra binding energy to nuclei. Thus, a ridge is 
seen for all the isotopic chains, peaking between $N=132-140$, with its 
breadth increasing with every increase in the $Z$ value of a chain. Consequently,
the neutron drip line is stretched from $N=152$ for Sm ($Z=62$) to far away 
$N=184$ for Hf ($Z=72$). Occasional hiccups in the ridges arise due 
to a shape transition and/or to a change in magnitude of deformation 
as compared to a neighbouring nucleus.

\begin{figure}[t]
\resizebox{0.40\textwidth}{!}{%
   \includegraphics{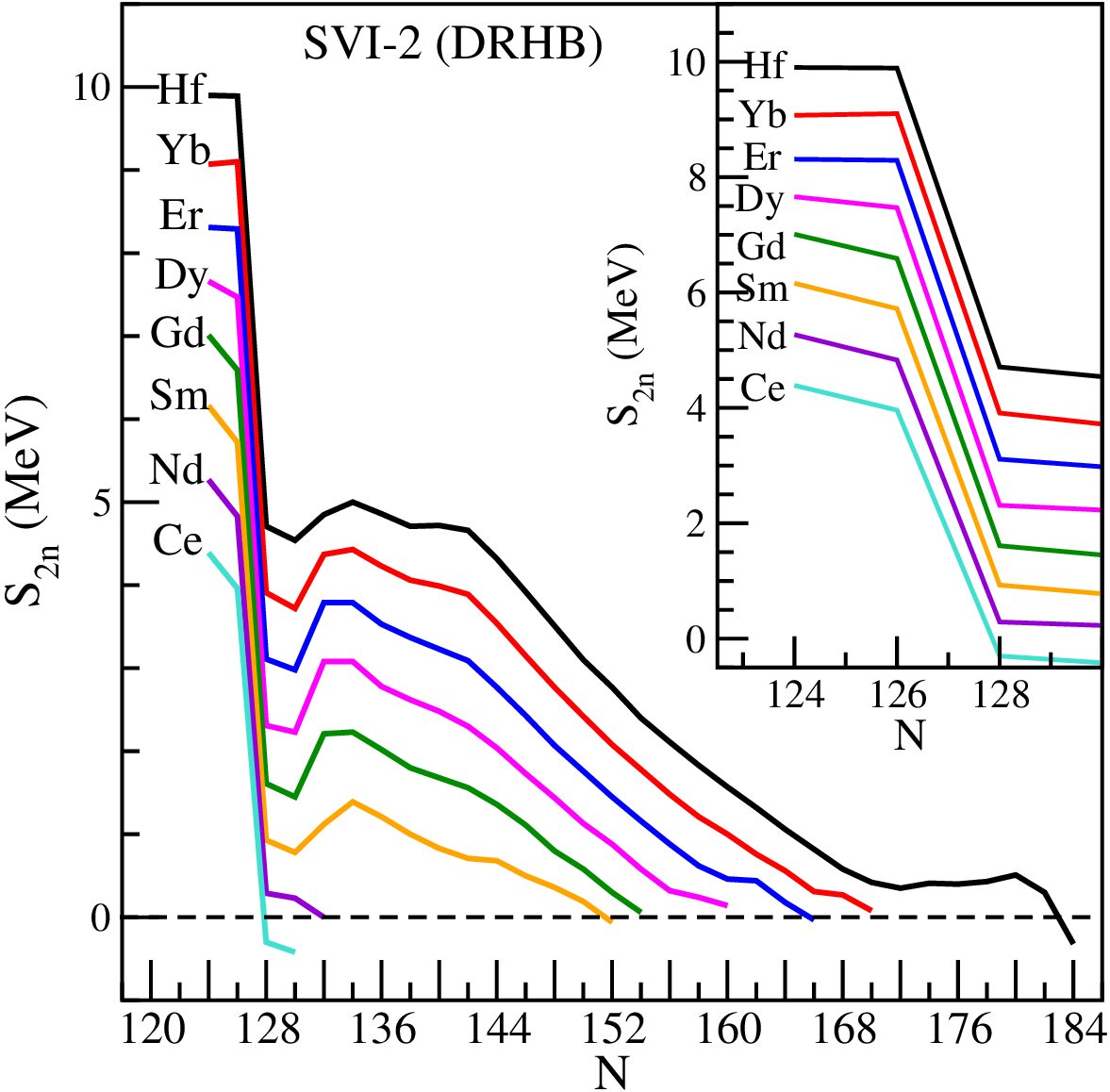}}
\caption{The same as in Fig.~\ref{fig:fig13} but for the 
Lagrangian model SVI-2.}
\label{fig:fig14}       
\end{figure}

The $S_{2n}$ values obtained with the Lagrangian model SVI-2 are shown in 
Fig.~\ref{fig:fig14}. The behavior and the features of $S_{2n}$ values
with SVI-2 are also very similar to those with NL-SV1 (Fig.~\ref{fig:fig13}).
The shell effects at $N=126$ at the neutron
drip line with SVI-2 are also strong as seen from the inset of 
Fig.~\ref{fig:fig14}, including for the Ce chain. The pronounced ridge
at $N \approx 134$ and a secondary ridge at $N \approx 142$ for Hf ($Z=72$),
Yb ($Z=70$) and Er ($Z=68$) are also seen with SVI-2. The region of extra
stability in nuclei and the extension of the region of bound nuclei 
from $N=128$ to $N=182$ for Hf chain is similar to that with NL-SV1.
In contrast, the Nd chain does not show the ridge, for
the neutron drip line is arrived imminently at $N=128$.
The tail of each curve shows that the neutron drip line with 
SVI-2 for various chains especially the heavier elements occurs 
earlier at about 2-4 neutrons less than that with NL-SV1.

\begin{figure}[t]
\hspace{-0.5 cm}
\resizebox{0.40\textwidth}{!}{%
   \includegraphics{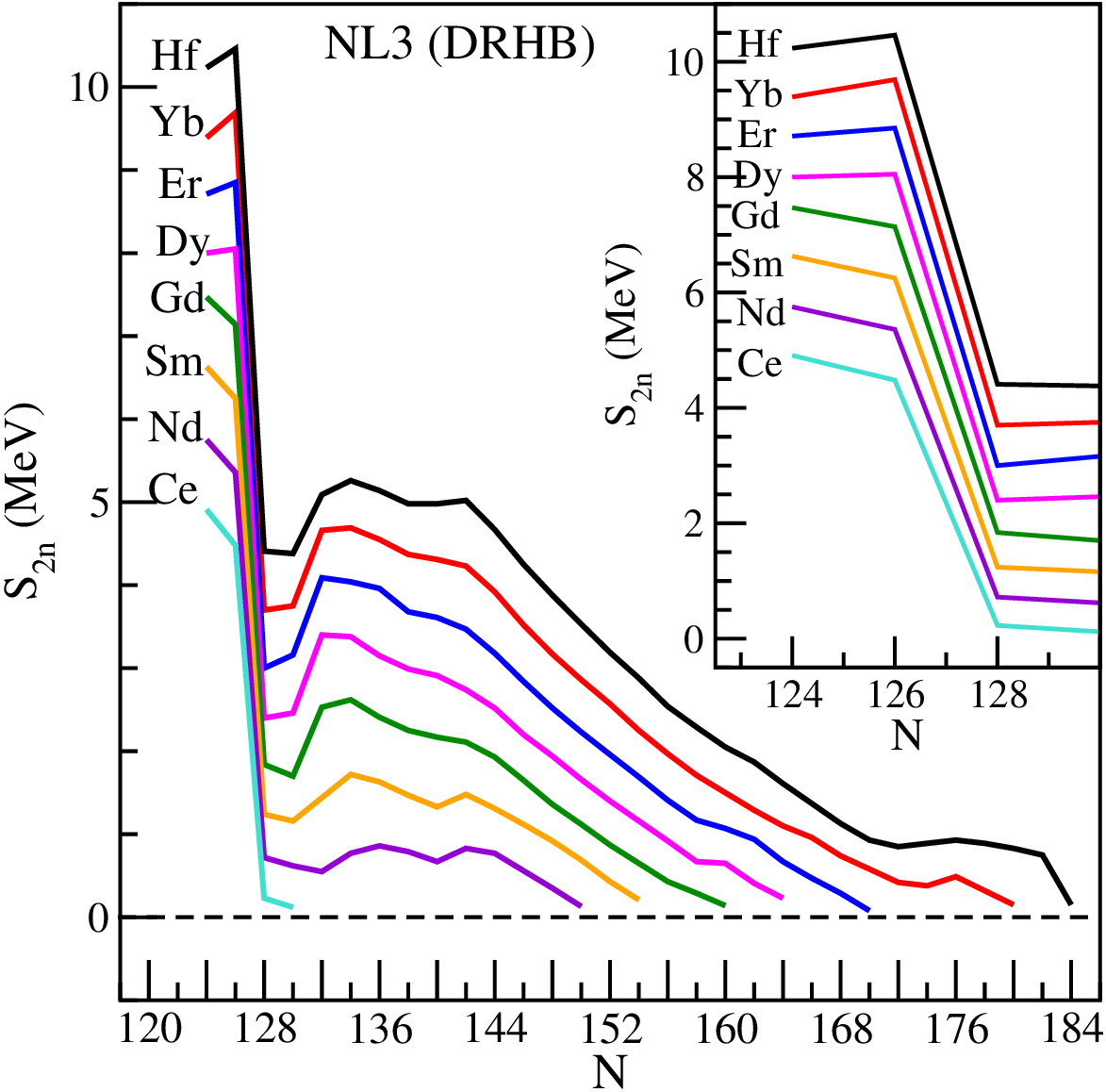}}
\caption{The same as in Fig.~\ref{fig:fig13} but for the set NL3 of the  
Lagrangian model NL-$\sigma$.}
\label{fig:fig15}       
\end{figure}

The $S_{2n}$ values obtained from DRHB calculations using the Lagrangian 
model NL3 are shown in Fig.~\ref{fig:fig15}. The $S_{2n}$ curves for
various isotopic chains are very similar to those seen with NL-SV1 in 
Fig.~\ref{fig:fig13} and with SVI-2 in Fig.~\ref{fig:fig14}. 
A marginal decrease in shell effects at $N=126$ 
in going from Hf ($Z=72$) to Ce ($Z=58$) near the neutron drip line is
seen in the inset of Fig.~\ref{fig:fig15}. However, the shell
effects at the magic number $N=126$ remain robust near the drip line
with NL3 as well. The creation of ridges of stability engendered
by deformation is very similar to that seen with NL-SV1 ans SVI-2. 
Except for the Nd ($N=60$) chain, where the neutron drip line
is seen to occur at $N=150$ as against at $N=144$ for NL-SV1, 
the neutron drip line with NL3 is seen to be stretched from
$N=154$ for Sm ($Z=62$) to $N=184$ for Hf ($Z=72$). Thus,
the neutron drip line as seen in this figure for NL3 occurs with
about 2-4 neutrons more than that with NL-SV1, and 4-6 neutrons
more than that with SVI-2. This is ostensibly due to
an overbinding of nuclei with NL3, as seen in 
Figs.~\ref{fig:fig2}-\ref{fig:fig4}.

A remark about a small ridge near the end of the $S_{2n}$ curves for
Hf ($Z=72$) isotopic chain in Figs.~\ref{fig:fig13}-\ref{fig:fig15} 
is in order. A falling $S_{2n}$ curve at $N\approx168$ for Hf 
must mean an imminent arrival of the drip line.
However, a maintenance of still significant deformation continues
until $N=180$ arrives. The deformation keeps adding binding energy to nuclei
above $N=170$ and a fast reduction of $S_{2n}$ to a zero value is averted. This leads to a
nearly flat region of $S_{2n}$ between $N=170$ and 180 with a little ridge
created in the region. Thus, due to this effect, the drip line for
Hf is extended up to $N \approx 184$ for all the three Lagrangian models. 
This effect is visible to a little extent in a small peak at $N=176$ for 
Yb ($Z=70$) with NL3 in Fig.~\ref{fig:fig12}. The overbinding by NL3
in the extreme regions has extended the drip line for Yb to $N=180$
as compared to that at $N=172$ for NL-SV1 (Fig.~\ref{fig:fig13}) and
at $N=170$ for SVI-2 (Fig.~\ref{fig:fig14}).   

\subsection{The shell gap at $N=126$}

The difference between $S_{2n}$ value at the magic number and that for
a nucleus with two neutrons above is a good reflection of the magnitude of the 
shell gap $\Delta$, as given by
\begin{equation}
\label{eq:delta}
\Delta (N_0) =  S_{2n} (N_0) - S_{2n} (N_0+2). 
\end{equation}
As seen in the inset of Fig.~\ref{fig:fig13}, the shell gap at $N=126$
shows only a marginal decrease in going down from Hf ($Z=72$) to Ce ($Z=58$).
The $S_{2n}$ value for Ce becomes zero at $N=128$, immediately above the shell
closure, pointing to an arrival of the neutron drip line just above $N=126$.
Yet the shell gap at $N=126$ for Ce remains large even in the midst of 
the neutron drip line. This implies that there is no weakening (quenching) 
of the shell effects at $N=126$ even as the neutron drip line approaches.
Thus, the shell effects at $N=126$ near the neutron drip line remain 
overly strong. It confirms the strong shell effects near the neutron drip line
predicted within the framework of the spherical relativistic Hartree-Bogoliubov 
approach \cite{FS.06}.  

\begin{figure}[h]
\resizebox{0.40\textwidth}{!}{%
   \includegraphics{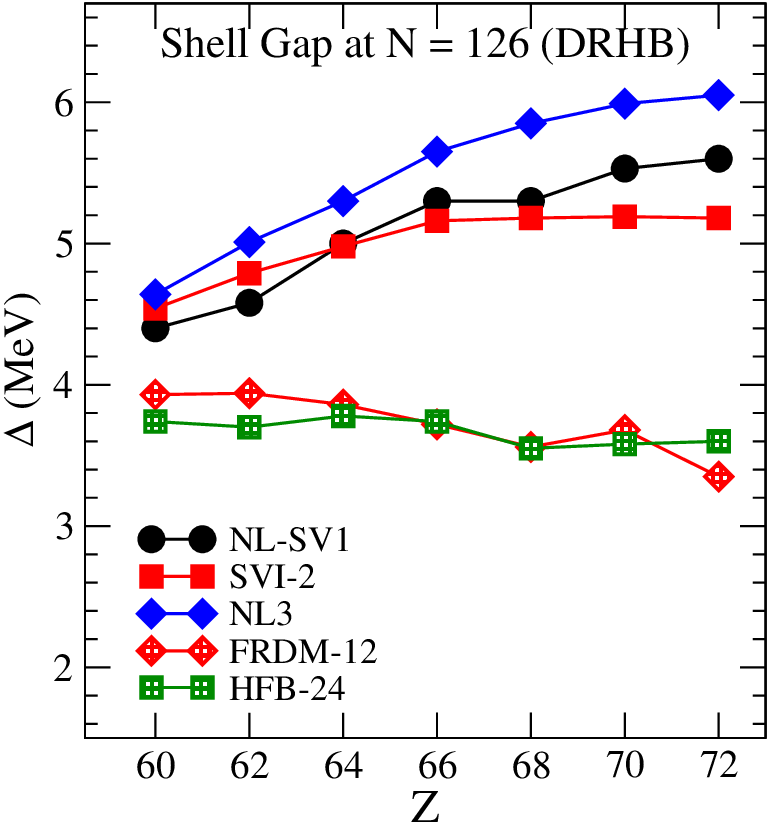}}
\caption{The neutron shell gap $\Delta$ at the magic number
$N=126$ for various isotopic chains obtained with the Lagrangian models NL-SV1, 
NL3 and SVI-2. A comparison is made with the shell gaps from the predictions 
of the mass formulas FRDM-12 and HFB-24.}
\label{fig:fig16}       
\end{figure}

The shell gap $\Delta$ calculated  from $S_{2n}$ values for various Lagrangian
models are shown in Fig.~\ref{fig:fig16}. As noted in the insets of 
Figs.~\ref{fig:fig13}-\ref{fig:fig15} for change of
$S_{2n}$ values across the shell closure at $N=126$, the shell gap $\Delta$
shows a steady but marginal decline in going from Hf ($Z=72$) to Nd ($Z=60$).
The shell gap is more spread for Hf and the heavier chains amongst
various Lagrangian models. NL3 shows the largest shell gap whereas SVI-2 does
the least. Predictions of NL-SV1 and SVI-2 on binding energies and deformations
as seen before have in general been closer to each other. This is also 
reflected in the shell gaps shown here. For isotopes below Yb ($Z=70$), the 
shell gaps with NL-SV1 and SVI-2 merge with each other. For the extreme
cases of Sm and Nd which are closer to the neutron dril line, the shell
gaps with NL3 also merge with those of NL-SV1 and SVI-2. 

Shell gaps with SVI-2, in particular, remain constant at $\approx$ 5.0 MeV in going 
from Hf ($Z=72$) to Dy ($Z=66$). These show a small decline to $\approx$ 4.5 MeV in going
to the Nd ($Z=60$) near the neutron drip line. The results show that the 
shell effects with the RMF Lagrangian models remain robust even in the
vicinity of the neutron drip line.

The shell gaps from predictions of the mass formulas FRDM-12 and HFB-24
are also shown for comparison in Fig.~\ref{fig:fig16}. Interestingly,
shell gaps with both the mass formulas show nearly identical behaviour.
The shell gaps at $N=126$ with the mass formulas are in general much lower than
those with the RMF models especially in the region of Gd ($Z=64$) to Hf ($Z=72$), 
which are quite far away from the drip line.
Shell gaps with the mass formulas are $\approx 3.5$ MeV for Hf ($Z=72$), Yb ($Z=70$) 
and Er ($Z=66$). In comparison, the shell gaps with SVI-2 are $\approx$ 50\% larger 
than those of the mass formulas and the shell gaps with NL3 are 
about $\approx$ 70\% larger than those of the mass formulas.

Whilst the shell gaps with HFB-24 are nearly constant over the whole range of
$Z$ values, with a marginal increase in going towards the drip line ($Z=60$),
FRDM-12 shows a tendency of a slight increase towards the drip line. This
behaviour of the mass formulas especially that of FRDM-24 is contrasting
with that of the RMF models which show a decrease in going towards the neutron
drip line. Near constancy of the shell gaps with the mass formulas may be due
to 'structure-less' shell structure. On the other hand, the shell gaps with the mass
formulas and the RMF models in going towards the neutron drip line at $N=126$
approach each other. The differences between the two in the region of the drip
line ($Z=60-62$) are small ($\approx$0.5 MeV). Nonetheless,  the robustness
of the shell effects at $N=126$ close to the neutron drip line in the 
RMF models is consistent with the predictions of the mass formulas. 

\begin{figure}
\hspace {-0.5 cm}
\resizebox{0.45\textwidth}{!}{%
   \includegraphics{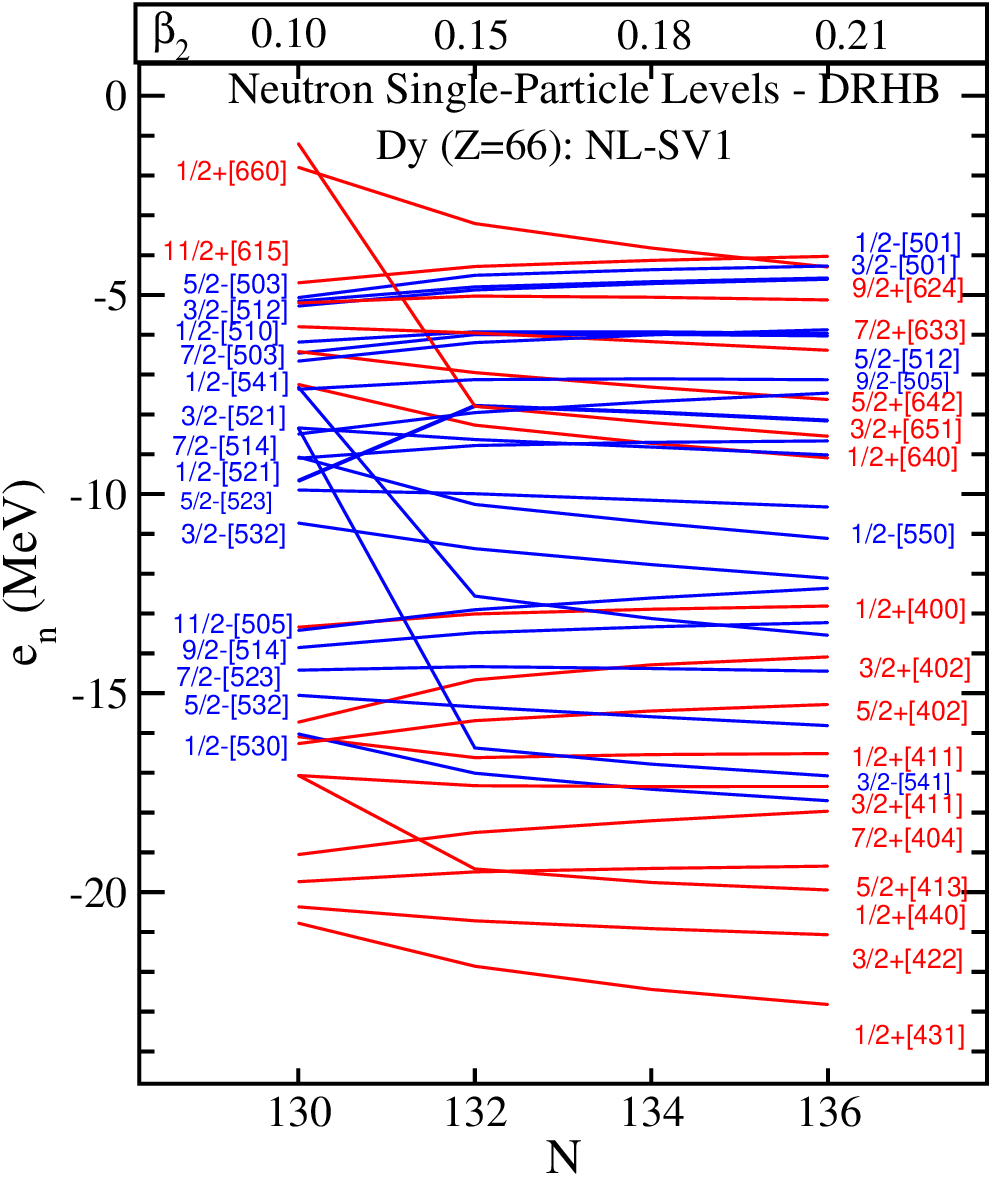}}
\caption{The neutron single-particle (Nilsson) levels 
$\Omega^\pi[N,n_s,\Lambda]$ for the isotopes
of Dy ($Z=66$) obtained with NL-SV1. The positive parity
levels are shown by red (in coloured figures) and the negative parity ones
by blue (in coloured figures). The $\beta_2$ values of the nuclides 
are shown at the top of the panel.}
\label{fig:fig17}       
\end{figure}
\subsection{Single-particle levels in nuclei}

With a view to visualize as to how the enhanced stability beyond a
presumed neutron drip line arises, the single-particle levels (Nilsson)
$[N,n_s,\Lambda]\Omega^\pi$ of nuclei are examined.
The neutron single-particle levels for some isotopes of Dy ($Z=66$)
are shown in Fig.~\ref{fig:fig17} as obtained with NL-SV1.
Dy chain is taken as a representative to illustrate the structure.

\begin{figure}
\resizebox{0.45\textwidth}{!}{%
   \includegraphics{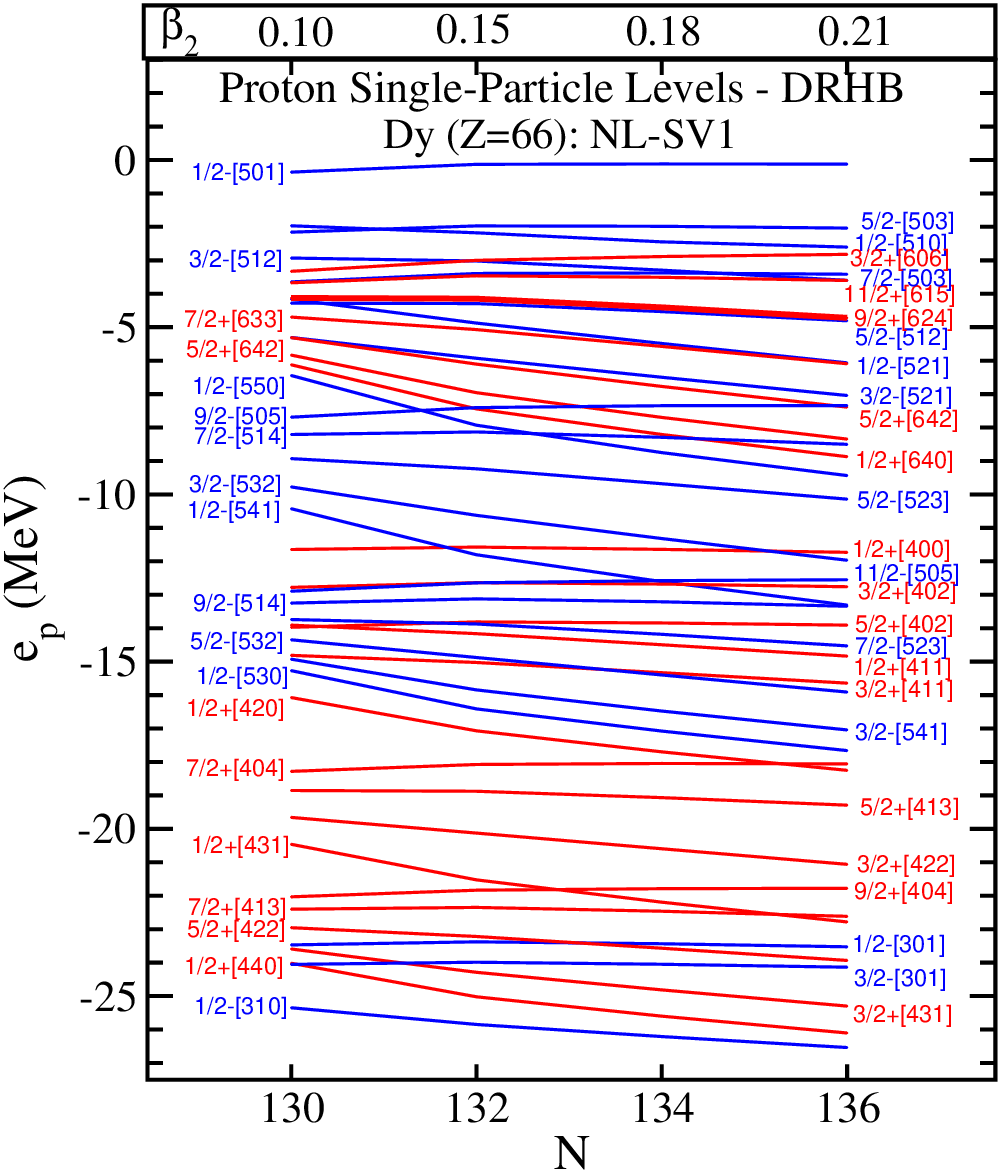}}
\vspace{0.4 cm}
\caption{The same for protons as in Fig.~\ref{fig:fig17}.} 
\label{fig:fig18}       
\end{figure}
The isotopes $N=124-128$ are all spherical. As an onset of deformation 
begins at $N=130$, the nuclei become deformed with 
$\beta_2 \approx 0.10-0.21$. The effect of this deformation is 
evident in the splitting of $j$-levels into various 
$\Omega^\pi$-orbitals. It is well known that for higher shells, relatively
small deformations produce large changes in energy. Deformation
introduces a change in the Nilsson Hamiltonian as given by
\begin{equation}
\label{eq:nilsson}
H' = -M\omega_0^2r^2\delta.\frac{2}{3}
\sqrt{\frac{4\pi}{5}}Y_{20}(\theta,\phi)
\end{equation}
where deformation $\delta$ is written in terms of the deformation
parameter $\beta_2$ as
\begin{equation}
\label{eq:def}
\beta_2 = \frac{2}{3}\sqrt{\frac{4\pi}{5}}\delta.
\end{equation}

Heavy nuclei are more susceptible to a gain in energy in passing 
from a spherical to a deformed shape \cite{deShalit.74}.
Particles that extend to larger distances (large $l$) are  
strongly affected by a deformation that may even be modest 
(Eq. \ref{eq:nilsson}). This feature of the deformed single-particle 
levels is clearly visible in Fig.~\ref{fig:fig17} where some of 
the $\Omega^\pi$-levels are pushed down significantly in energy. 

Both positive and negative parity levels are pushed down 
in energy at and beyond $N=132$, especially those with the lowest
value of $\Omega^\pi = 1/2$ as expected for a prolate deformation.
The positive parity level ${3/2}^+[651]$ is more prominently pushed 
down whereas ${1/2}^+[660]$, ${1/2}^+[440]$ and some other 
levels do so modestly. The level ${3/2}^+[651]$, a component of 
the single-particle state $1i_{13/2}$, that is lying very high 
(and is with a high $l$) at $N=130$ with only $37\%$ occupancy is 
pushed down significantly and is filled up
when $N=132$ is encountered. A few other positive parity levels especially
with higher $\Omega^\pi$ values are marginally lifted up, with 
a little consequence on the total binding energy.

A major contribution to an increase in binding energies 
of nuclei with $N \ge 132$ arises from a significant 
push down of the high-lying $-$ve parity level ${3/2}^-[541]$. 
This is a part of the spatially outer lying single-particle 
level $1h_{11/2}$. The $-$ve parity levels ${1/2}^-[541]$ and ${1/2}^-[521]$,
with the former going down and the latter going up at $N \ge 132$, 
are a component of the neighbouring single-particle levels 
$2f_{7/2}$ and $3p_{3/2}$, respectively.

The net effect of many levels going down and a few going up is 
to effectively lower the 'centre-of-gravity' of the 
neutron single-particle levels. This is then responsible for 
the additional binding energy provided by the deformation to 
nuclei beyond the expected neutron drip line. The same effect is 
visible in all the isotopic chains showing an onset of
a ridge in $S_{2n}$ values at $N > 130$. In heavier isotopic chains,
the ridge in $S_{2n}$ values is extended well up until $N \approx 150$. Herein,
a push-down of higher $l$ and low $\Omega^\pi$ levels especially 
$\Omega = 1/2$ of $1i_{13/2}, 1i_{11/2}, 1h_{11/2}, 1h_{9/2}, 2g_{9/2}$
and $2g_{7/2}$ orbitals would contribute to additional binding energies arising from 
the deformation.

The corresponding proton single-particle levels are shown
in Fig.~\ref{fig:fig18}. The proton single-particles with
a fixed number of protons for a given isotopic chain are not
affected much by the onset of deformation with only a
marginal lowering of the energy levels. It is not expected
that the proton single-particle levels would be abruptly
lowered down by an onset of deformation on addition of
more neutrons.   

\section{\label{sec:conclude}Summary and conclusions}

The structure of nuclei beyond $N=126$, which are relevant to the
third peak ($A \approx 195$) in the $r$-process nucleosynthesis is investigated
using three different Lagrangian models. The  nuclei spanned are from the
neutron drip line at $N=126$ to the neutron drip line near
$N=184$. The study has brought out several interesting features:

1. The three Lagrangian models show an agreement on binding energy of nuclei
within a few MeV, for nuclei that are not too far away from $N=126$. This
is especially true for the heavier rare-earth chains ($Z > 64$).
Close to the the neutron drip line near $N=126$, the differences amongst
the RMF models increase. The binding energies due to the models NL-SV1
($NL-\sigma\omega$), and SVI-2 (Model SVI) are closer to each other. However,
the results with NL3 (NL-$\sigma$) diverge the most from the other
two models. NL3 overestimates the binding energies by upto $\approx$10 MeV as
compared to NL-SV1 and SVI-2. For the lighter chains of Gd ($Z=64$ and Sm
($Z=62$), divergences amongst various models increase especially
near the drip line. The results due to NL-SV1 and SVI-2 are still closer
to each other.  

2. A comparison of the binding energies for nuclei in heavier chains
($Z >62$) shows interesting results. Comparison of the results from the RMF
models with those from the semi-empirical mass formulas FRDM-2012 and HBF-24
shows that FRDM-2012
underbinds nuclei as compared to HFB-24 moderately and but significantly as
compared to the RMF models. This is especially the case near the drip
line. Notably, the binding energies due to SVI-2 show very good
agreement with the predictions from HFB-24 for various chains but
especially so for the heavier chains. In comparison, NL3
shows the largest deviations from HFB-24 and FRDM-2012. For instance,
the $rms$ deviation from the HFB values for Er ($Z=68$) is 0.7 MeV
for SVI-2 and is 2.1 MeV for NL-SV1 (see Table II). The $rms$
deviation for NL3 is 6.1 MeV. Similar results are seen for the other chains
(see Table III for Yb, $Z=70$). It turns out that the Lagrangian model SVI-2
produces results that are qualitatively similar to those of the much
improvised mass formula HFB-24. The Lagrangian model SVI-2 without inclusion
of non-linearities of $\sigma$- and $\omega$-mesons may be employed to construct a
mass formula. However, it needs to be investigated in other regions of the
periodic table as well. A comparison of the RMF result with those of the mass
formula DRHBc shows that the latter overesitnates the binding energies of
nuclei rather excessively, reflected by an $rms$-deviation of 11.6 MeV for the Er
(Z=68) chain, for instance. This feature is seen for all the isotope chains
studied here.

3. The strength of the shell effects at $N=126$ in the neutron-rich region
especially close to the neutron drip-line has been a point of conjecture in
the absence of reasonable data in this region. The mass formulas FRDM-2012 and
HFB-24 and others predict strong shell effects at $N=126$. Our investigation
shows that all the RMF models predict strong shell effects on going
near to the neutron drip-line. The strength of the shell effects can
have a decisive influence on abundances around the third peak at $A \approx 190$,
as seen in a recent work with network chain calculations \cite{Li.2026}.

4. An onset of deformation takes place for most of the chains ($Z \ge 62$), that
begins at $N \approx 134$ with nuclei assuming a prolate shape. The deformation
peaks at around $\beta_2 \approx 0.25 - 0.30$ as more neutrons are added. A shape
transition from a prolate shape to an oblate shape takes place for all chains
except for Hf ($Z=72$). The point of transition shifts to higher neutron
numbers as the $Z$ value of a chain increases. This is seen with all the Lagrangian
models (Fig.~\ref{fig:fig5} - Fig.~\ref{fig:fig7}).

5. The onset of deformation and the increase of deformation to
nuclei near the mid-shell leads to a preponderance of shape-coexistence
of a prolate and oblate shape in many chains from Sm ($Z=6$2) to Hf ($Z=72$).
This is observed with all the Lagrangian models. In the heavier chains,
a significant number of nuclei exhibit a
shape-coexistence between a prolate shape and an oblate shape in the ground state
akin to a near degeneracy. This effect is seen with all the Lagrangian models.
Needless to say that such an effect can hardly be predicted within the
semi-empirical mass formulas. In any case, this phenomenon is expected to
influence the nuclear abundances due to different half-lives for different
shapes. 

6. Nuclei acquiring significantly large deformation create a
region of enhanced stability. The region of bound nuclei increases
significantly in the form of ridges of stability depicted
by $S_{2n}$  for chains above Sm ($Z=62$). Consequently, the
$r$-process path is dramatically shifted to higher neutron numbers as one goes
from Sm ($Z=62$) to Hf ($Z=72$). The neutron drip line is shifted from
$N \approx 150$ for Sm  to very near the magic number $N=184$ for Hf. This is
symptomatic of all the Lagrangian models investigated here.
The increased stability rendered to a large number of nuclei in various chains
is due to lowering down of several neutron single-particle levels
$\Omega^\pi$ levels in energy (see Fig.~\ref{fig:fig17} and Fig.~\ref{fig:fig18}). 

7. The ground state of nuclei near the $r$-process path and the neutron
drip-line are predicted to be oblate for most of the isotopic chains studied
in this work. Nuclei with oblate shape would have longer $\beta$-decay
half-lives. This would imply a slight resistance to nucleosynthesis.
However, an admixture of nearly prolate shape in coexistence with an oblate
shape would accelerate the nucleosythesis. A combination of these factors
would affect the abundance curve near the third peak significantly.
Reaction network calculations using $\beta$-decay rates obtained from
deformation and shape coexistence should cast a realistic picture of
$r$-process nuclear abundances.

\begin{acknowledgments}
 The author acknowledges useful discussions with Dr. Ameena R. Farhan. The
 author thanks Peter Ring for the program on DRHB. The author is thankful to
 the referee for pointing out the the recent mass formula DRHBc that
 has brought to fore the constrasting differences amongst variuous models.
 His sagacity to suggest amendments to improve the presentation is gratefully
 acknowledged.
\end{acknowledgments}


\begin{thebibliography}{00}

\bibitem{BBFH.57} E.M. Burbidge {\it et al.}, Rev. Mod. Phys. {\textbf 29} 
(1957) 547.

\bibitem{Hille.78} W. Hillebrandt, Space Sci. Rev. {\textbf 21} (1978) 639.

\bibitem{Kratz.93} K.L. Kratz {\it et al.}, Astrophys. J. {\textbf 403} (1993) 216.

\bibitem{GW17.2017} B.P.Abbott {]it et al.} Phys. Rev. Lett. 
  {\textbf 119} (2017) 161101.

\bibitem{Thiel.2017} F.-K. Thielemann, M. Eichler, I.V. Panos, and
  B. Wehmeyer, Ann. Rev. Nucl. Part. Sci. {\textbf 67} (2017] 253.  

\bibitem{Pfeiffer.01} B. Pfeiffer {\it et al.}, Nucl. Phys. A {\textbf 693} 
(2001) 282.

\bibitem{Pfeiffer.97} B. Pfeiffer, K.L. Kratz, and F.K. Thielemann, 
Z. Phys. {\textbf A357} (1997) 235.

\bibitem{Dillmann.02} I. Dillmann {\it et al.}, 
Phys Rev. Lett. {\textbf 91} (2002) 162503.

\bibitem{Jungclaus.07} A. Jungclaus {\it et al.}, Phys. Rev. Lett. 
{\bf 99} (2007) 132501.

\bibitem{Rodriguez.08}  T.R. Rodriguez,  J.L. Egido and A. Jungclaus,
Phys. Lett. B {\bf 668} (2008) 410.

\bibitem{Gorska.09} M. Gorska {\it et al.}, Phys. Lett. B {\bf 672} (2009)
  313.

\bibitem{Hwang.08} J.K. Hwang, A.V. Ramayya, and J.H. Hamilton, J. Phys. G:
  Nucl. Part. Phys. \textbf{35} (2008) 055102.

\bibitem{SF.02} M.M. Sharma and A.R. Farhan, Phys. Rev. C \textbf{65} (2002)
  044301.

\bibitem{Surman.97} R. Surman, J. Engel, J.R. Benett, and B.S. Meyer.
  Phys. Rev. Lett. {\bf 79} (1997) 1809.

\bibitem{SFM.00} M.M. Sharma, A.R. Farhan and  S. Mythili, 
Phys. Rev. C \textbf{61} (2000) 054306.


\bibitem{FS.06} A.R. Farhan and M.M. Sharma, Phys. Rev. {\bf C73} (2006)
  045803.

\bibitem{Guzman.15} R. Rodriguez-Guzman, L.M. Robledo, and M.M. Sharma, 
Eur. Phys. J. \textbf{A51} (2015) 73.

\bibitem{SW.86} B.D. Serot and J.D. Walecka, Adv. Nucl. 
Phys. {\textbf 16} (1986) 1.
 
\bibitem{Lala.01} G.A. Lalazissis, D. Vretenar, and P. Ring,  
Nucl. Phys. A {\textbf 679} (2001) 481.

\bibitem{Berger.84} J.F. Berger, M. Girod, and D. Gogny, 
  Nucl. Phys. A {\textbf 428} (1984) 32.

\bibitem{ozel.16} F. \"Ozel, P. Freire,
  Ann. Rev. Astron. Astrophys. \textbf{54} (2016) 401

\bibitem{Dworschak.08} Dworschak {\it et al.}, Phys. Rev. Lett. {\textbf 100} (2008)
  072501. 

 \bibitem{Sharma.08} M.M. Sharma, Phys. Lett. B {\textbf 666} (2008) 140.

\bibitem{Fomenko.93} V.N. Fomenko, S. Marcos, L.N. Savushkin, 
  J. Phys. G: Nucl. Part. Phys. {\textbf 19} (1993) 545.

\bibitem{BB.77} J. Boguta, A.R. Bodmer, Nucl. Phys.{\textbf A292} (1977) 413.

\bibitem{GRT.90}Y.K. Gambhir, P. Ring, A. Thimet, Ann. Phys. (N.Y.) 
                {\textbf 198} (1990) 132.

\bibitem{Ring.96} P. Ring, Prog. Part. Nucl. Phys. \textbf{37} (1996) 193.
                
\bibitem{Rein.89} P.G. Reinhard, Rep. Prog. Phys. \textbf{52} (1989) 439.

\bibitem{Lee.86} S.J. Lee, J. Fink, A.B. Balantekin, M.R. Strayer, 
A.S. Umar, P.G. Reinhard, J.A. Maruhn, W. Greiner, 
Phys. Rev. Lett. \textbf{57} (1986) 2916.

\bibitem{SNR.93} M.M. Sharma, M.A. Nagarajan, P. Ring, Phys. Lett.
B \textbf{312} (1993) 377.

\bibitem{Lala.97} G.A. Lalazissis, J. K\"onig, and P. Ring, Phys. Rev. C
  \textbf{55} (1997) 540. 

\bibitem{Meng.2010} P.W. Zhao, Z.P. Li, J.M. Yao, and J. Meng, Phys. Rev. C
  \textbf{82} (2010) 054319.

\bibitem{Moeller.16} P. M\"oller, A.J. Sierk, T. Ichikawa, and H. Sagawa,
  At. Data Nucl. Data Tables \textbf{109-110} (1026) 1.

\bibitem{Goriely.13} S. Goriely, N. Chamel, and J.M. Pearson, Phys. Rev. C
  \textbf{88} (2013) 024308. 

\bibitem{Guo.2024} P. Guo et al., At. Data Nucl. Data Tables \textbf{158} 101661.

\bibitem{deShalit.74} A. deShalit and H. Feshbach, in {\it Theoretical Nuclear 
Physics, Vol. I: Nuclear Structure}, (1974) John Wiley \& Sons, Inc.
  
\bibitem{Li.2026} M. Li, G.C. McLaughlin, and R. Surman, Phys. Lett. B {\textbf 872} (2026) 140109.


\end{thebibliography}
\end{document}